\documentclass[journal]{IEEEtran}
\IEEEoverridecommandlockouts
\usepackage{algorithm} 
\usepackage{algpseudocode}
\usepackage{dsfont}
\usepackage{array}
\usepackage{hyperref}
\usepackage{cite}
\usepackage{tikz}
\usepackage{graphicx}
\usetikzlibrary{positioning}
\usepackage{listings}
\usepackage{amsmath,amssymb,amsfonts}
\usepackage{amsthm}
\usepackage{graphicx}
\usepackage{textcomp}
\usepackage{stmaryrd}
\usepackage{bbm}
\usepackage{bm}
\usepackage{soul}
\usepackage{amsmath}
\usepackage[utf8]{inputenc}
\usepackage[english]{babel}
\usepackage{mathtools} 
\usepackage{comment}
\usepackage{xcolor}

\usepackage{babel}

{}
\newtheorem{definition}{Definition}

\usepackage{caption}
\usepackage{subcaption}
\usepackage{xcolor}
\usepackage{todonotes}

\definecolor{codegreen}{rgb}{0,0.6,0}
\definecolor{codegray}{rgb}{0.5,0.5,0.5}
\definecolor{codepurple}{rgb}{0.58,0,0.82}
\definecolor{backcolour}{rgb}{0.95,0.95,0.92}

\lstdefinestyle{mystyle}{
    backgroundcolor=\color{backcolour},   
    commentstyle=\color{codegreen},
    keywordstyle=\color{blac},
    numberstyle=\tiny\color{codegray},
    stringstyle=\color{codepurple},
    basicstyle=\ttfamily\footnotesize,
    breakatwhitespace=false,         
    breaklines=true,                 
    captionpos=b,                    
    keepspaces=true,                 
    numbers=left,                    
    numbersep=5pt,                  
    showspaces=false,                
    showstringspaces=false,
    showtabs=false,                  
    tabsize=2
}
\allowdisplaybreaks
\usepackage{cite}
\usepackage{amsmath,amssymb,amsfonts}
\usepackage{graphicx}
\usepackage{textcomp}
\usepackage{xcolor}
\usepackage{capt-of}
\usepackage{dsfont}
\usepackage{epsfig,latexsym,amsmath,amssymb,cite,color,graphicx,cases}
\usepackage{amsmath,amssymb,amsfonts,cite}
\usepackage{amsthm}
\usepackage{graphicx}
\usepackage[percent]{overpic}
\usepackage{stmaryrd}
\usepackage{bbm}
\usepackage{bm}
\usepackage{amsmath}
\usepackage[utf8]{inputenc}
\usepackage[english]{babel}

\title{Helper-Assisted Coding for Gaussian Wiretap Channels: Deep Learning Meets PhySec}

\author{Vidhi Rana, R\'{e}mi A. Chou, Taejoon Kim  \thanks{V. Rana and R. Chou are with the Department of Computer Science and Engineering, University of Texas at Arlington, Arlington, TX, and T. Kim is with the School of Electrical, Computer and Energy Engineering, Arizona State University,
Tempe, AZ. E-mails: vidhi.rana@uta.edu, remi.chou@uta.edu, taejoonkim@asu.edu. Part of this work has been presented at the 2024 IEEE International Conference on Communications (ICC) \cite{rana2024helper}. This work was supported in part by National Science Foundation (NSF) under grants CNS2451268, CNS2514415, and the NSF and Office of the Under Secretary of Defense (OUSD) – Research and Engineering, Grant ITE2515378, as part of the NSF Convergence Accelerator Track G: Securely Operating Through 5G Infrastructure Program.}}

\begin{document}
\maketitle

\begin{abstract}
Consider the Gaussian wiretap channel, where a  transmitter wishes to send a confidential message to a legitimate receiver in the presence of an eavesdropper. It is well known that if the eavesdropper experiences less channel noise than the legitimate receiver, then it is impossible for the transmitter to achieve positive secrecy rates. A known solution to this issue consists in involving a second transmitter, referred to as a helper, to help the first transmitter to achieve security.
While such a solution has been studied for the asymptotic blocklength regime and via non-constructive coding schemes, in this paper, for the first time, we design explicit and short blocklength codes using deep learning and cryptographic tools to demonstrate the benefit and practicality of cooperation between two transmitters over the wiretap channel. Specifically, our proposed codes show strict improvement in terms of information leakage compared to existing codes that do not consider a helper. Our code design approach relies on a reliability layer, implemented with an autoencoder architecture based on the successive interference cancellation method, and a security layer implemented with universal hash functions. We also propose an alternative autoencoder architecture that significantly reduces training time by allowing the decoders to independently estimate messages without successively canceling interference by the receiver during training.  
Additionally, we show that our code design is also applicable to the multiple access wiretap channel with helpers, where two  transmitters send confidential messages to the legitimate receiver.

\end{abstract}


\section{Introduction}
Physical-layer security exploits physical characteristics of wireless channels to transmit confidential messages. An information-theoretic approach to physical-layer security has been first proposed by Wyner with the wiretap channel model in \cite{wyner}, where a transmitter wishes to send a confidential message $S$ to a legitimate receiver, Bob, in the presence of an eavesdropper, Eve. Bob's estimate of $S$ from his channel output observations is denoted by~$\widehat{S}$, and Eve’s channel output observations are denoted by
$Z^n$, where $n$ is the number of channel uses. In \cite{wyner}, the constraints are that Bob must be able to recover $S$, i.e., $\lim_{n\rightarrow \infty}\mathbb{P}[S\neq \widehat{S}]=0$, and the leakage about $S$ at Eve must vanish with $n$, i.e., $\lim_{n\rightarrow \infty}\frac{1}{n}I(S;Z^n)=0$. Note that the stronger security requirement $\lim_{n\rightarrow \infty}I(S;Z^n)=0$
 has also been  considered~\cite{maurer}, meaning that Eve's observations~$Z^n$ are almost independent of $S$ for large $n$. The maximum transmission rate for which Bob can reliably decode the transmitted message and secrecy is guaranteed over a wiretap channel is called secrecy capacity. Unfortunately, if the channel conditions are such that the channel gain between the transmitter and the eavesdropper is better than the channel gain between the transmitter and the legitimate receiver, then the secrecy capacity is zero, meaning that it is impossible for the transmitter to send a secret message to the legitimate receiver~\cite{wyner,cheong}.  

To overcome this impossibility, settings that involve multiple users need to be considered to go beyond point-to-point transmission, and corresponding codes need to be designed for such settings. The works \cite{tekin, xie2012secure,he2014providing,chou2022gaussian} demonstrate that cooperation between transmitters can be beneficial to enable positive secrecy rates at the transmitters who could not achieve positive secrecy rates with point-to-point codes. For instance, the Gaussian wiretap channel in the presence of a helper is the extension of the Gaussian wiretap channel, where a helper cooperates with the transmitter such that 
the interferences between their codewords prevent the eavesdropper from retrieving the confidential message sent by the transmitter.

While existing works have mainly focused on the asymptotic blocklength regime and non-constructive coding schemes to derive achievability secrecy rates, in this paper, we propose to design explicit and short blocklength codes {\color{black}($\leq 64$)} for the Gaussian wiretap channel in the presence of a helper that cooperates with the transmitter under information-theoretic security guarantees. Specifically, security is quantified as information leakage at the eavesdropper which is measured in terms of the mutual information between the confidential message and the eavesdropper's channel observations. Our constructed  codes demonstrate the benefit of user cooperation over a wiretap channel at short blocklength as follows:
\begin{itemize}
\item We show that for a transmitter with adverse channel conditions (i.e., his channel gain with the eavesdropper is better than his channel gain with the legitimate receiver), a second transmitter, called helper, can help decrease information leakage at the eavesdropper. Note that such a result is impossible to achieve by solely relying on point-to-point Gaussian wiretap channel codes, and therefore, new codes that enable cooperation need to be designed. 
\item We show that for a transmitter with favorable channel conditions (i.e., his channel gain with the eavesdropper is worse than his channel gain with the legitimate receiver), a helper can also help decrease information leakage at the eavesdropper.
\end{itemize}

  {Our main contributions are as follows:
 \begin{enumerate}
     \item \textcolor{black}{  We propose the first construction of finite blocklength codes for the Gaussian wiretap channel that supports the presence of a helper, applicable to blocklengths up to $n\leq 64$  -- designing codes for
larger blocklengths remains a challenging open problem.} Our proposed construction consists of one reliability layer and one security layer. The reliability layer is implemented with an autoencoder architecture inspired by the 
    successive interference cancellation (SIC) idea, first introduced for broadcast channels in \cite{cover1972broadcast}, and the security layer is implemented with universal hash functions. The performance evaluation of our proposed codes show significant improvement in information leakage in comparison to best-known codes  that do not consider a~helper (referred to as point-to-point codes).

     \item We propose an alternative autoencoder architecture for the design of the reliability layer, where the receiver estimates the messages without canceling the interference during training. As seen in our simulations, this alternative architecture reduces the training time compared to the architecture based on the
     SIC method. Using this approach, we design codes to support multiple helpers, and we have shown significant improvement in information leakage with multiple helpers.

     \item We demonstrate that our proposed code design is also suitable for the Gaussian multiple access wiretap channel, where two legitimate transmitters communicate with a legitimate receiver, as demonstrated by our simulations results in Section \ref{mac_h_sim}. Additionally, our simulation results for this setting are consistent with previous work that considered non-constructive coding schemes and the asymptotic regime, e.g., \cite{tekin}, where a helper can help decrease the information leakage at the eavesdropper.
     \item  We propose a modular code design that consists of two layers: a \textcolor{black}{security layer} and a reliability layer. The \textcolor{black}{security layer} only addresses the secrecy constraint and depends solely on the statistics of the eavesdropper’s channel. In contrast, the reliability layer only addresses the reliability constraint and depends solely on the statistics of the legitimate receiver’s channel. This approach allows a simplified code design. Additionally, through the independent design of two layers, we can precisely control the level of information leakage at the eavesdropper. 
 \end{enumerate}}
\textcolor{black}{Note that finite-length achievability and converse bounds on the secrecy rate for the Gaussian wiretap channel have only been derived in the absence of a   helper in \cite{yang}, and consequently do not apply to our setting. Moreover, even   without security constraint,  such as in a multiple-access channel, obtaining non-asymptotic achievability and converse bounds beyond a second-order achievable rate \cite{molavianjazi2009arbitrary} remains challenging. This main roadblock is our motivation to study from a practical point of view the design of codes for cooperating users over a Gaussian wiretap channel.}
 
The remainder of the paper is organized as follows. Section~\ref{rev} reviews related works. Section~\ref{model_des} introduces the Gaussian wiretap channel model with a helper. Section~\ref{jm_sc} describes our proposed code design and its performance evaluation for the Gaussian wiretap channel model with one helper.  Section~\ref{gwtc_arch2} proposes a code design for the Gaussian wiretap channel model to support multiple helpers. Additionally, we present an alternative autoencoder architecture that simplifies the decoder, leading to reduced training time.  Section~\ref{mac_h} discusses the Gaussian multiple access wiretap channel and presents our simulation results. Finally, Section~\ref{cr} provides concluding remarks. 
 \section{Related works}\label{rev}
To the best of our knowledge, prior to this work, no finite-length code constructions have been proposed for cooperating users over a wiretap channel under an information-theoretic leakage security metric.  
We discuss next related works that have studied  the asymptotic,   i.e., the regime where blocklength approaches infinity, and non-asymptotic blocklength regimes.
\subsection{Works on the asymptotic blocklength regime}   


\textcolor{black}{For the asymptotic blocklength regime, the works \cite{tekin,xie2012secure,he2014providing,chou2022gaussian} have already demonstrated that cooperation between transmitters can be beneficial to enable positive secrecy rates at the transmitters who could not achieve positive secrecy rates with point-to-point codes.
 In \cite{tekin}, \cite[Theorem 1]{tekin} shows that in multiple access scenarios,  cooperative users help to enable positive secrecy rates that would not be possible in a single-transmitter scenario. \cite{he2014providing} analyzes a Gaussian wiretap channel with a cooperative jammer, demonstrating that the secure degrees of freedom (d.o.f.), defined as $\lim_{P \rightarrow \infty}\frac{Cs}{\frac{1}{2}\log P}$ with transmit power $P$ and secrecy capacity $C_s$, can be positive. Without the jammer, $C_s$ does not scale with $P$, yielding zero secure d.o.f.. \cite{xie2012secure}  studies a Gaussian wiretap channel with $L$ helpers and show that the secure d.o.f. of the Gaussian wiretap channel with helpers when transmitting power goes to infinity is $\frac{L}{L+1}$.    \cite{chou2022gaussian} studies a Gaussian multiple-access wiretap channel in the presence of a jammer-aided eavesdropper and  \cite[Theorem 2]{chou2022gaussian}  achieves  positive secrecy rates that would not be possible  without cooperation. Note also that \cite{tekin,xie2012secure,he2014providing,chou2022gaussian} focus on providing non-constructive schemes through  random coding arguments that do not provide any guidelines to construct coding schemes. In contrast, constructive coding schemes exist for the asymptotic blocklength regime, such as low-density parity-check (LDPC) codes for the two-way Gaussian wiretap channel \cite{pierrot_ldpc} and polar codes for the multiple access wiretap channel \cite{chou2018}. 
 }

\subsection{Works on the non-asymptotic blocklength regime}
\subsubsection{Coding theory}
There are no prior finite-length code constructions for multiple transmitters   over a wiretap channel.
The most related works at finite blocklengths are for single-transmitter (point-to-point) wiretap channels under (i) 
non-information-theoretic security metrics and (ii) an
information-theoretic security metric.  Coding-theoretic constructions based on the non-information-theoretic security metric called security gap, which is based on an error probability analysis at the eavesdropper, 
include randomized convolutional codes for Gaussian and binary symmetric wiretap channels \cite{r2}, randomized turbo codes for the Gaussian wiretap channel~\cite{noora}, and LDPC codes for the Gaussian wiretap channel \cite{klinc,baldi}. Additionally, another non-information-theoretic security approach called practical secrecy is studied in~\cite{harri}, where a leakage between Alice's message and an estimate of the message at Eve is estimated. 
Coding-theoretic constructions based on an information-theoretic security metric include punctured systematic irregular LDPC codes  \cite{r4},  LDPC codes \cite{baldi2},  and randomized Reed-Muller codes \cite{r1} for the Gaussian wiretap channel. 


\subsubsection{Deep learning}

Machine learning approaches for channel coding where secrecy constraints are not taken into account include deep learning \cite{oshea, dorner, r3, kim2023}, reinforcement learning \cite{aoudia, goutay}, and generative adversarial~networks \cite{ye2018channel}. In \cite{oshea}, a setting with two transmitters is also investigated.

Recently, deep learning approaches for channel coding have been extended to wiretap channel coding in the presence of a \emph{single transmitter}. In \cite{gtc_frit,dl_frit}, a coding scheme that imitates coset coding by clustering learned signal constellations is developed for the Gaussian wiretap channel under a secrecy metric that relies on a cross-entropy loss function. 
In \cite{besser}, a coding scheme for the Gaussian wiretap channel is developed under  an information-theoretic security metric whose approach consists in training an autoencoder to optimize the reliability and secrecy constraints simultaneously, and  \cite{rana2023short,rana2021design,sultana2023secret}, whose approach aims to separately handle the reliability constraint via an autoencoder and the secrecy constraint via hash functions for better modularity and a fine information leakage control.   Unlike \cite{rana2023short,rana2021design,sultana2023secret} that consider a single transmitter, we consider multiple transmitters in this paper, which can be seen as multiple interfering autoencoders trying to estimate messages, with each autoencoder having a different objective. This setting becomes more challenging to train as the number of training parameters grows with the number of transmitters, and novel training architectures for code designs are needed to overcome this challenge. 
\paragraph*{Notation} We use uppercase for random variables and corresponding lowercase for their realizations, e.g., $x$ is a realization of random variable $X$.  We also write $\vert \mathcal{X} \vert$ to represent the cardinality of a set $ \mathcal{X}$. {\color{black}Let $\mathbb{B}^n_0(r)$ be the ball of radius $r$ centered at the origin in $\mathbb{R}^n$ under the Euclidian~norm.}

\section{Model}\label{model_des}
\begin{figure}[h]
\centering
\includegraphics[trim=7.0cm 4cm 8cm 5cm,clip,width=0.52\textwidth]{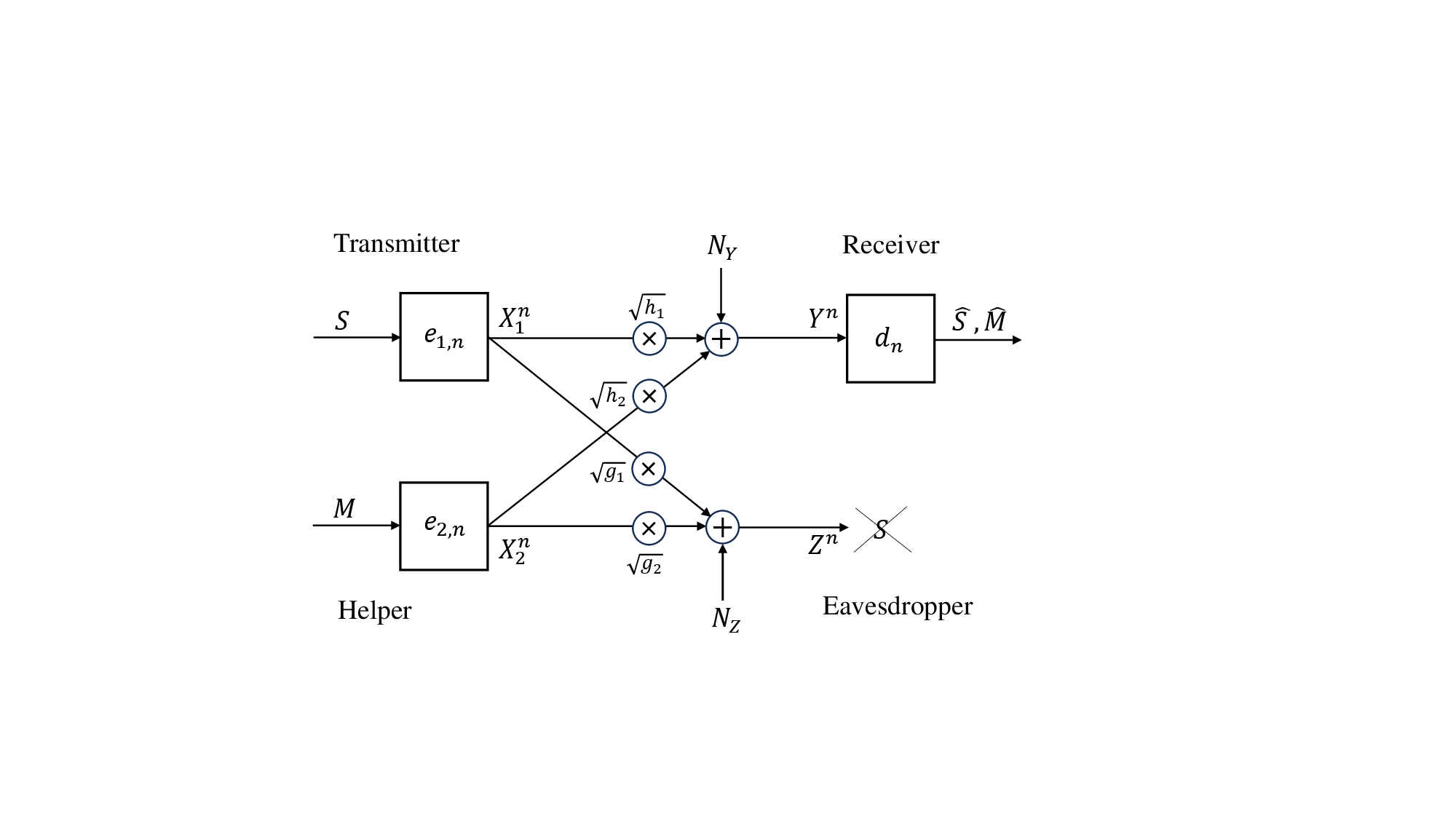}
\caption{ Gaussian wiretap channel with a helper.}
\label{GWTC_H}
\end{figure}

As depicted in Figure \ref{GWTC_H}, we consider a Gaussian wiretap channel with a helper defined by
\begin{align}
    Y&\triangleq\sqrt{h_1}X_1+ \sqrt{h_2}X_2+N_Y\label{mod1a},\\
    Z&\triangleq\sqrt{g_1}X_1+\sqrt{g_2}X_2+N_Z\label{mod1b},
\end{align}
where $h_1$ and $h_2$ are the channel gains of the transmitter and helper, respectively, to the intended receiver, $g_1$ and $g_2$ are the channel gains of the transmitter and helper, respectively, to the eavesdropper, and $N_Y$ and $N_Z$ are zero-mean Gaussian random variables with variances $\sigma^2_Y$ and $\sigma^2_Z$, respectively. The legitimate receiver and the eavesdropper observe the sequences $Y^n$  and $Z^n$, respectively, obtained from (\ref{mod1a}) and (\ref{mod1b}), and all the above channel parameters are known to everyone. 

In this model, the transmitter wishes to transmit a secret message $S$ to the legitimate receiver by cooperating with the helper who wishes to transmit an unprotected message $M$ to the legitimate~receiver.

\begin{definition}
 A $(k_1,k_2, n,P_1, P_2)$  Gaussian  wiretap channel code with a helper consists of
\begin{itemize}
 
        \item an encoder for the transmitter
        $$e_{1,n}:\{0,1\}^{k_1}\rightarrow \mathbb{B}^n_0(\sqrt{nP_1}),$$ which, for a message $S\in \{0,1\}^{k_1}$, forms the codeword~$X_1^n\triangleq~e_{1,n}(S)$;
         \item an encoder  for the helper
          $$e_{2,n}:\{0,1\}^{k_2}\rightarrow \mathbb{B}^n_0(\sqrt{nP_2}),$$ which, for a message $M\in \{0,1\}^{k_2}$, forms the codeword~$X_2^n\triangleq~e_{2,n}(M)$;
          
        \item a decoder for the legitimate receiver \\
        $$d_{n}: \mathbb{R}^n\rightarrow  \{0,1\}^{k_1} \times \{0,1\}^{k_2},$$ which, from the channel observations $Y^n$, forms an estimate of the messages $(S,M)$ as  $(\widehat{S},\widehat{M})\triangleq d_{n}(Y^n)$\textcolor{black}{.}
  
\end{itemize}

The codomain of the encoders reflects the following power constraints for the transmitter and helper 
\begin{align}
    \sum_{t=1}^{n}(X_i(t))^2\leq nP_i,~i\in\{1,2\}, \label{eqPw}
\end{align}
where $X_i(t)$ is the $t$-th entry of $X_i^n$. 
\end{definition}

The performance of a $( k_1,k_2,n,P_1, P_2)$ code is measured in terms of
\begin{enumerate}
    \item The average probability of error for the secret message~$S$ \begin{align}
   \mathbf{P}^{(S)}_e\triangleq \frac{1}{2^{k_1}}\sum^{2^{k_1}}_{s=1}\mathbb{P}[ \widehat{S} \neq s\vert s~\text{is~sent} ];  \label{eqpe1}    
    \end{align}
    \item The average probability of error for the unprotected message $M$ \begin{align}\mathbf{P}^{(M)}_e\triangleq \frac{1}{2^{k_2}}\sum^{2^{k_2}}_{m=1}\mathbb{P}[ \widehat{M}\neq m\vert m~\text{is~sent} ]; \label{eqpe2}  
    \end{align}
    \item The information leakage about the message $S$ at the eavesdropper \begin{align}\mathbf{L}_e\triangleq {I}(S;Z^n). \label{eql}\end{align}
\end{enumerate}

\begin{definition}
A $(k_1,k_2, n,P_1, P_2)$ code is said to be $(\epsilon_S,\epsilon_M)$-reliable if $\mathbf{P}_e^{(S)}\leq \epsilon_S$ and $\mathbf{P}_e^{(M)}\leq \epsilon_M$,  and $\delta$-secure if $\mathbf{L_e}\leq \delta$. Moreover, a rate pair $(\tfrac{k_1}{n},\tfrac{k_2}{n})$ is  $(\epsilon_S,\epsilon_M,\delta)$-achievable with power constraint $(P_1,P_2)$ if there exists an $(\epsilon_S,\epsilon_M)$-reliable and $\delta$-secure $(k_1,k_2, n,P_1, P_2)$ code.
  \end{definition}

Note that the encoders $(e_{1,n},e_{2,n})$ and the decoder $d_n$ are public knowledge and known to the eavesdropper.

\section{Coding Scheme based on SIC}\label{jm_sc}
We first describe, at a high level, our coding scheme in Section \ref{cs_hld_case1}. Our coding approach consists of two coding layers: a reliability layer, whose design is described in Section~\ref {rellyr}, and a security layer, whose design is described in Section~\ref{SL}. 
Finally, we provide simulation results and examples of our code designs in Section \ref{sim_fdts}. Our simulation results show a significant advantage in terms of information leakage having a helper compared to having no helper.

\subsection{High-level description of our coding scheme }\label{cs_hld_case1}

\begin{figure}[ht]
\centering
\includegraphics[trim=6.5cm 1.1cm 1.8cm 3.8cm,clip,width=0.65\textwidth]{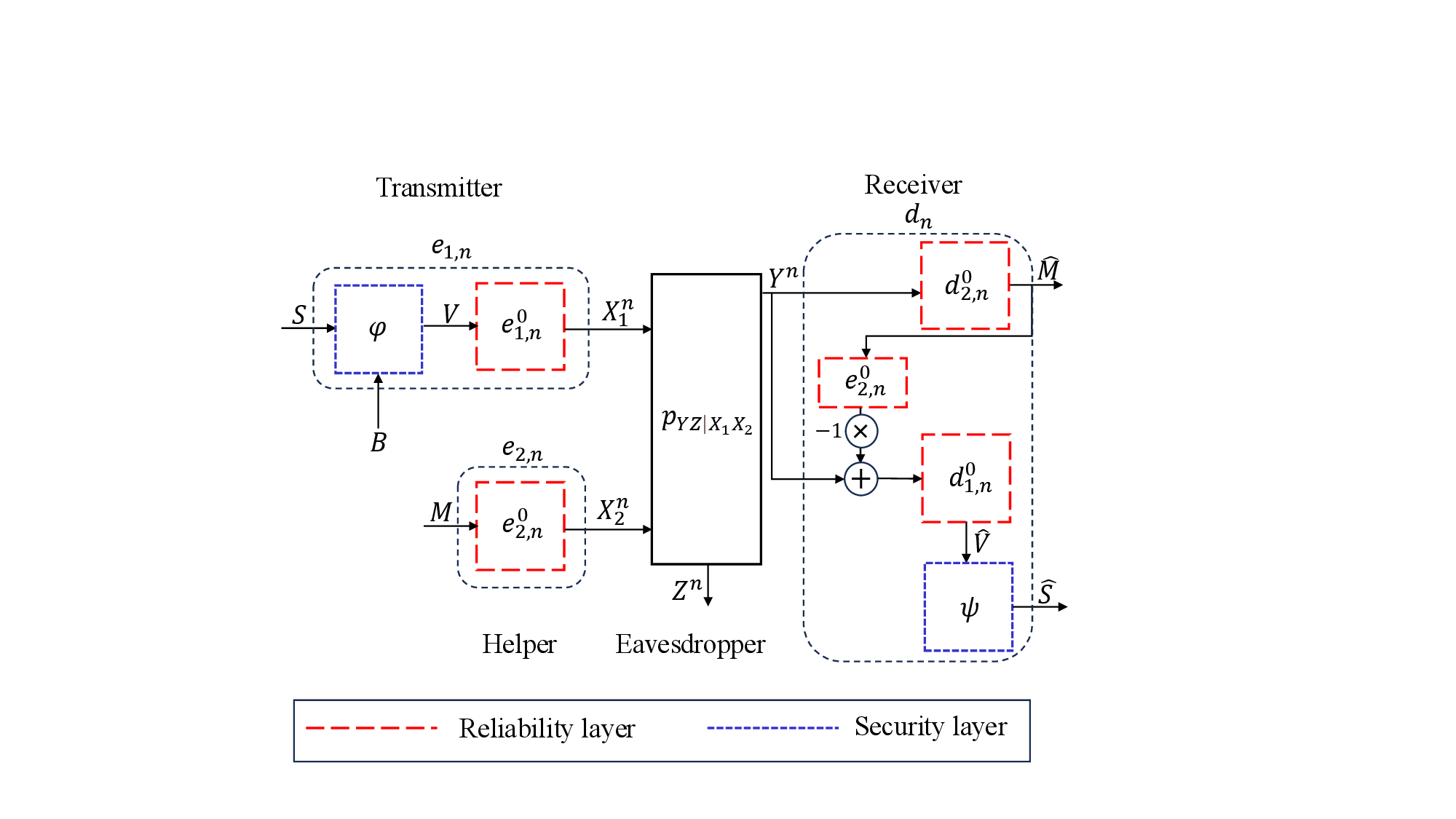}
\caption{Our code design consists of a reliability layer and a security layer. The reliability layer is implemented using two encoders $(e^0_{1,n}, e^0_{2,n})$ and two decoders $(d^0_{1, n}, d^0_{2, n})$, and the security layer is implemented using the functions $\varphi$ and $\psi$. }
\label{mac}
\end{figure}
As shown in Figure \ref{mac}, our code construction consists of (i) a reliability layer implemented with an $(\epsilon_S, \epsilon_M)$-reliable  $(q_1, q_2, n, P_1, P_2)$ code described by the encoders $e^0_{1,n}$ for the transmitter and $e^0_{2,n}$ for the helper, and a decoder pair $(d^0_{1,n},d^0_{2,n})$ for the legitimate receiver,\footnote{This code is designed without any security requirement, i.e., its performance is solely measured in terms of the average probability of errors \eqref{eqpe1},~\eqref{eqpe2}.}  and (ii) a security layer for the transmitter, which consists of an encoding function  $\varphi$ and a decoding function $\psi$. Note that for the helper there is no \textcolor{black}{security layer}, as the security constraint \eqref{eql}  concerns only the transmitter. As detailed in Sections \ref{rellyr} and \ref{SL}, we will design the reliability layer using a deep learning approach based on a neural network autoencoder with SIC, {\color{black}where we assume $h_2P_2 \geq h_1P_1$, such that $M$ is decoded first, followed by $V$ (if $h_2P_2 < h_1P_1$ the decoding order is reversed),}  
and the \textcolor{black}{security layer} using universal hash functions. 

{\it Encoding at the transmitter}:   
The transmitter first generates a sequence $B$ of $q_1-k_1$ bits uniformly at random in~$\{0,1\}^{q_1-k_1}$ with $q_1\geq k_1$, which represents local randomness used to randomize the output of the function $\varphi$ to confuse the eavesdropper. Then, the transmitter encodes the message $S$ uniformly distributed in $\{0,1\}^{k_1}$ as $e^0_{1,n}(\varphi(S, B))$. The overall encoding map $e_{1,n}$ for the transmitter that describes both secrecy and reliability layers is described by  
\begin{align*}
    e_{1,n}:\{0,1\}^{k_1}\times \{0,1\}^{q_1-k_1}& \rightarrow{}\mathbb{B}^n_0(\sqrt{nP_1}),\\ 
   (s,b) &\mapsto e^0_{1,n}(\varphi(s,b)).
\end{align*}

{\it Encoding at the helper}: The helper encodes the message $M$ uniformly distributed in $\{0,1\}^{k_2}$ as $e^0_{2,n}(M)$.

{\it Decoding}: From the channel observations $Y^n$, the legitimate receiver successively decodes $M$ and $S$. Specifically, the message $M$ is first decoded as $\widehat{M} \triangleq d^0_{2,n}(Y^n)$, and then  the message $S$ is decoded as $\psi(d^0_{1,n}(Y^n-\sqrt{h_2}\widehat{X}_2^n))$, where  $\widehat{X}_2^n\triangleq e^0_{2,n}(\widehat{M})$.

\subsection{Design of the reliability layer}\label{rellyr}

\begin{figure}[ht]
\centering
\includegraphics[trim=4.8cm 0cm 2cm 0cm,clip,width=0.55\textwidth]{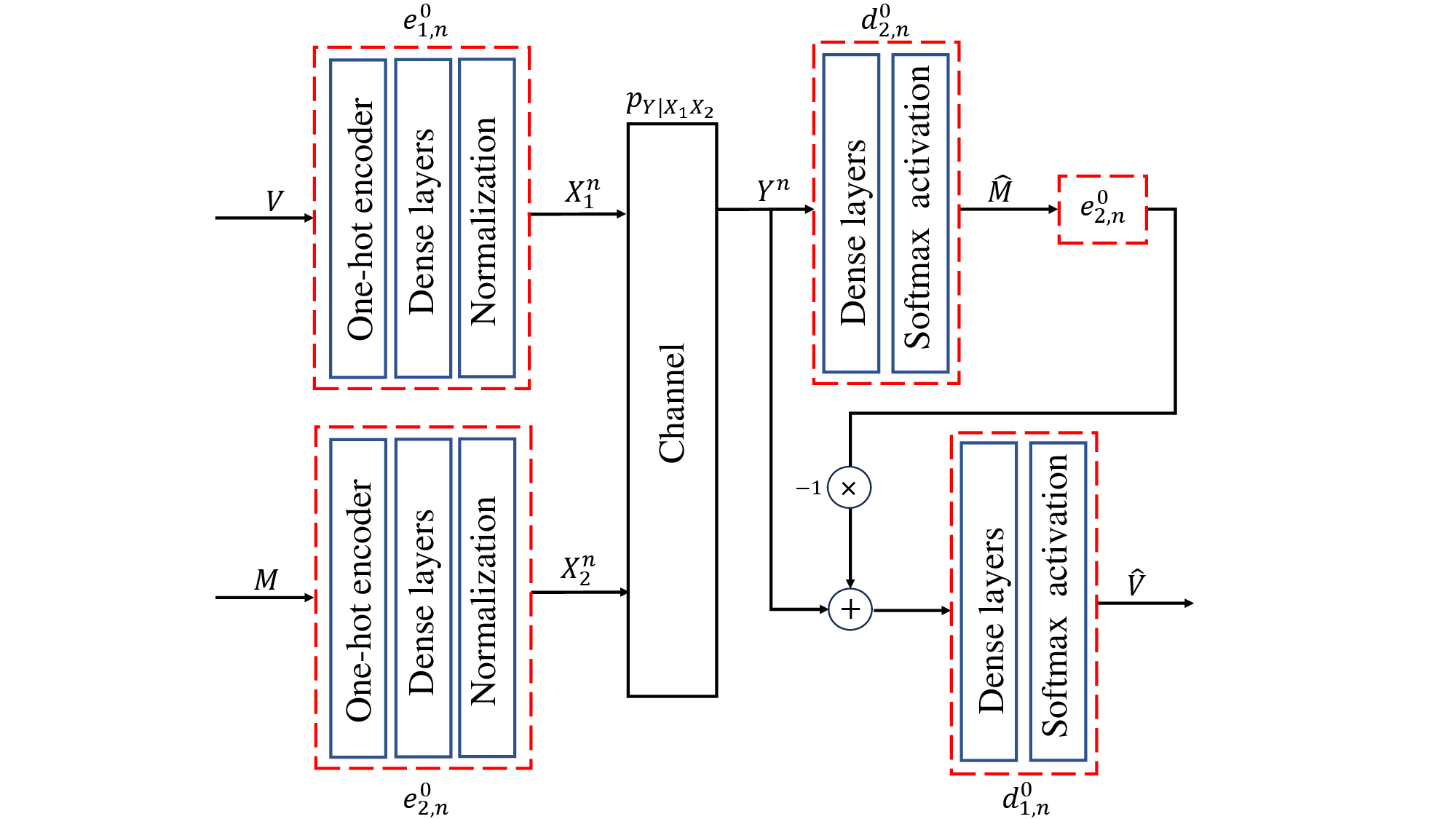}

\caption{Architecture of the autoencoder based on SIC.} \label{MAC_RL}
\end{figure}

The design of the reliability layer consists in designing an $(\epsilon_S,\epsilon_M)$-reliable $(q_1, q_2, n, P_1, P_2)$ code described by two encoders $(e^0_{1,n}, e^0_{2,n})$ and two decoders $(d^0_{1, n}, d^0_{2, n})$ for the channel described by \eqref{mod1a}.  
The input message $V \in \mathcal{V}\triangleq\{1,2,\dots, 2^{q_1}\}$ is encoded using a neural network encoder with an encoding function $e^0_{1,n}$ to obtain the codeword $X_1^n$, and the input message $M \in \mathcal{M}\triangleq\{1,2,\dots, 2^{q_2}\}$ is encoded using a neural network encoder with an encoding function $e^0_{2,n}$ to obtain the codeword $X_2^n$.
As depicted in Figure \ref{MAC_RL}, the encoders $e^0_{1,n}$ and $e^0_{2,n}$ consist of (i) an input layer where the message is fed to a one-hot encoder, which is followed by (ii)  dense layers with the ReLU activation function, followed by (iii) a dense layer that returns a vector of dimension $n$, followed by (iv) a normalization layer that ensures that the average power constraints \eqref{eqPw} are met for the codewords. The decoder  receives the channel output $Y^n$ and applies the decoder pair $(d^0_{1,n},d^0_{2,n})$ to successively estimate the messages ${M}$ and ${V}$, as shown in Figure \ref{MAC_RL},  which is inspired by the well-known SIC method, e.g., \cite{cover1972broadcast}. Specifically, upon receiving $Y^n$, the decoder $d^0_{2,n}$ recovers $M$ as $\widehat{M}$, while treating the signal $\sqrt{h_1}X^n_1$ from the transmitter as noise. Then, the receiver subtracts $\sqrt{h_2}X^n_2$ from $Y^n$ and the decoder $d^0_{1,n}$ decodes $V$ as $\widehat{V}$. 

As depicted in Figure \ref{MAC_RL}, the neural network decoders $(d^0_{1,n},d^0_{2,n})$ consist of dense layers with ReLU activation and a final layer with the softmax activation function whose output is a probability vector over all possible messages. Finally, the decoded messages correspond to the index associated with the highest probability. The autoencoder is trained over all
possible messages $v\in\mathcal{V}$ and $m\in \mathcal{M}$ using an \textcolor{black}{adaptive moment estimation (ADAM)} optimizer and the categorical cross-entropy loss function.  
\begin{figure}
      \centering
      \includegraphics[width=0.50\textwidth]{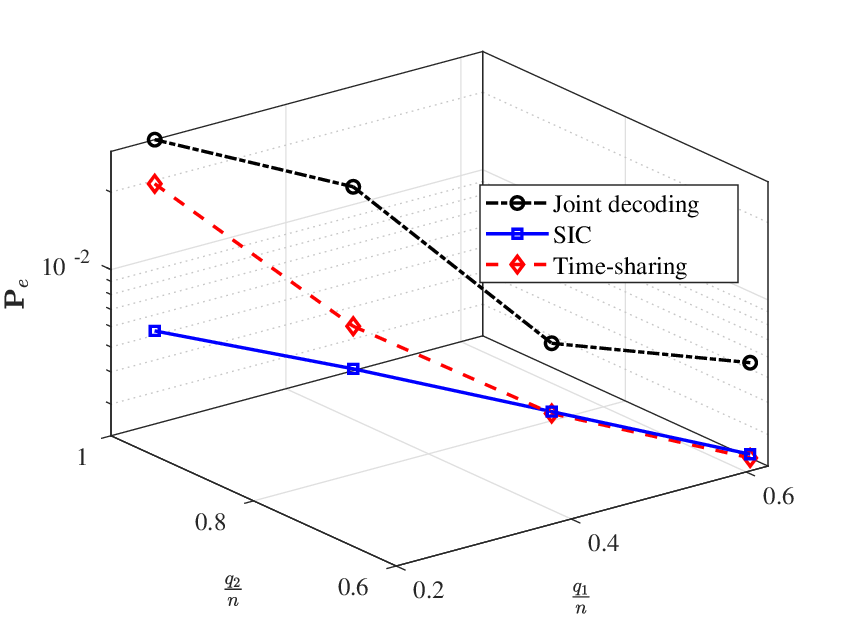}
      \caption{Comparison of schemes based on joint decoding, SIC, and time-sharing when $h_1=1$, $h_2=1$, $n=8$, $P_1=2$, $P_2=2$, and $\sigma^2_Y=6$.}
      \label{fig:comparisons}
\end{figure}
\paragraph*{Comparison between the SIC approach and other coding approaches}Figure \ref{fig:comparisons} compares the probability of error $\mathbf{P}_e \triangleq \mathbb{P}\left[(\widehat{V},\widehat{M})\neq (V,M)\right]$ of our proposed code design with (i) time sharing, and (ii) joint decoding. We observe that, depending on the rate pair $(\frac{q_1}{n}, \frac{q_2}{n})$, 
our proposed code design has similar or better performance, in terms of probability of error, than time-sharing.  We also observe that our approach outperforms joint decoding-based code designs, as the latter are difficult to train due to the complex decoder architecture that decodes messages simultaneously. For the time-sharing approach, we divided the time frame into two subframes, one of length $n_1$ and the other of length $n_2$, which are optimized to minimize the probability of error. During the first subframe, the transmitter encodes  $V$ as $X_1^{n_1}$, and the receiver observes $Y^{n_1}$ and decodes $V$ as $\widehat{V}$. During the second subframe, the helper encodes $M$ as $X_2^{n_2}$, and the receiver observes $Y^{n_2}$ and decodes $M$ as $\widehat{M}$. Note that for a fair comparison with our proposed code design, we chose the power constraints $P_1^{TS}=\frac{P_1}{\alpha}$ and $P_2^{TS}=\frac{P_2}{1-\alpha}$, for the transmitter and the helper, respectively, with  $\alpha \triangleq \frac{n_1}{n}$ and $n_1+n_2=n$. {\color{black}For the code design based on joint decoding, the neural network decoder consists of one decoder $d^0_n$ instead of two decoders $(d^0_{1,n},d^0_{2,n})$, i.e., the receiver simultaneously estimates $V$ and~$M$. The neural network decoder $d^0_n$ consists of $2^{q_1}+2^{q_2}$ dense layers with ReLU activation and a final layer with the softmax activation function whose output is a probability vector over all possible messages.}

  \subsection{Design of the security layer }\label{SL}  

 The objective of the \textcolor{black}{security layer} $(\varphi,\psi)$   is to limit the total amount of leaked information about the message $S$ in the sense that $I(S; Z^n)\leq \delta$, for some $\delta>0$.  To this end, we use 2-universal hash functions, whose definition is reviewed~next. 
 \begin{definition}\cite{carter} 
Given two finite sets $\mathcal{X}$ and $\mathcal{Y}$, a family~$\mathcal{G}$ of functions from $\mathcal{X}$ to $\mathcal{Y}$ is~2-universal if $ \forall x_1,x_2\in \mathcal{X},~x_1\neq x_2 \implies \mathbb{P}[G(x_1)=G(x_2)]\leq  \vert \mathcal{Y} \vert^{-1},
$ 
where $G$ is the random variable that represents the choice of a function $g\in \mathcal{G}$ uniformly at random in $\mathcal{G}$.
\end{definition}

Let $\Lambda\triangleq\{0,1\}^{q_1} \backslash \{\mathbf{0}\}$. For $k_1\leq q_1$, consider the  2-universal  family of hash functions $\mathcal{F}\triangleq( \psi_{\lambda})_{\lambda\in \Lambda}$, where for $ \lambda\in \Lambda$,  
\begin{align}\label{hash_r}
   \psi_{\lambda}:\{0,1\}^{q_1} &\rightarrow \{0,1\}^{k_1}, \\
    v &\mapsto(\lambda\odot v)_{k_1},
\end{align}
where $\odot$ is the multiplication in $\textup{GF}(2^{q_1})$ and $(\cdot)_{k_1}$ selects the $k_1$ left-most bits.  Then, we define
\begin{align}\label{hash_t}
 \varphi_{\lambda}: \{0,1\}^{k_1}\times \{0,1\}^{q_1-k_1} &\rightarrow \{0,1\}^{q_1}, \nonumber\\
    (s,b)& \mapsto \lambda^{-1}\odot(s \Vert b),
\end{align}
where $(\cdot\Vert\cdot)$ denotes the concatenation of two strings. Note that for any $\lambda \in \Lambda$, $s\in \{0,1\}^{k_1}$, $b\in \{0,1\}^{q_1-k_1}$, we have $\psi_{\lambda} \circ \varphi_{\lambda} (s,b) =s$.

In our proposed code construction, the design of the security layer consists in choosing the seed $\lambda\in \Lambda$ to minimize the information leakage at the eavesdropper. Additionally, the performance of the security layer will be evaluated using a mutual information neural estimator (MINE) \cite{r3a} and contrastive log-ratio upper bound (CLUB) \cite{cheng2020clubcontrastivelogratioupper} estimator. \textcolor{black}{While the mutual information neural estimator (MINE)  converges towards the true value of the mutual information $I(S;Z^{n})$ as the number of samples are sufficiently large\cite{r3a}, this limit is approached via a lower bound on mutual information. We further validated our results by estimating our leakage through the CLUB estimator, which   converges towards the true value of the mutual information via an upper bound on mutual~information.}

When the \textcolor{black}{security layer} is combined with the reliability layer, our coding scheme can be summarized as follows. The input of the encoder $e^0_{1,n}$ is obtained by computing $V\triangleq \varphi_{\lambda}(S, B)$, where $S\in \{0,1\}^{k_1}$ is the message, and $B\in \{0,1\}^{q_1-k_1}$ is a sequence of $q_1-k_1$ random bits generated uniformly at random. After computing $V$, the trained encoder $e_{1,n}^0$ generates the codeword $X_1^n\triangleq e_{1,n}^0(V)$. The input of the encoder $e^0_{2,n}$ is the message $M \in \{0,1\}^{q_2}$, and the output is the codeword of the helper $X_2^n\triangleq e_{2,n}^0(M)$. Then, the codewords $X_1^n$ and $X_2^n$ are sent over the channel, and the intended receiver and the wiretapper observe $Y^n$ and $Z^n$, respectively, as described by \eqref{mod1a} and \eqref{mod1b}. The legitimate receiver    decodes $Y^n$ as $\widehat{M} \triangleq d^0_{2,n}(Y^n)$ and $\widehat{V}\triangleq d^0_{1,n}(Y^n-\sqrt{h_2}\widehat{X}_2^n)$, where $\widehat{X}_2^n \triangleq e^0_{2,n}(\widehat{M})$. Finally, the receiver 
creates an estimate  $\widehat{S}$ of $ S$ as $\widehat{S}\triangleq \psi_{\lambda}( \widehat{V})$.

\subsection{Simulations}\label{sim_fdts}  
 We now provide examples of code designs that follow the guidelines described in Sections~\ref{rellyr}, \ref{SL}, and evaluate their performance in terms of the average probability of error at the receiver and information leakage at the eavesdropper. Note that no other finite-length codes have been proposed for our specific setting, therefore, our performance evaluation serves to quantify the gains introduced by the helper in terms of probability of error and information leakage when contrasted with a scenario where no helper is present. 
The neural networks are implemented in Python 3.8 using Tensorflow~2.6.2.
Based on the channel parameters in \eqref{mod1a} and \eqref{mod1b}, we consider two cases. 

\textbf{Case 1}: $\frac{h_1}{\sigma_Y^2}\leq \frac{g_1}{\sigma_Z^2}$. In this case, the eavesdropper has a channel advantage over the legitimate receiver. Intuitively, this means that the legitimate receiver experiences more channel noise than the eavesdropper does, and it is well known that the secrecy capacity is zero in this case \cite{cheong}. Therefore,  point-to-point codes cannot allow secure transmission of the message $S$ for the transmitter, and additional resources are needed to achieve security. Here, a helper represents such a resource that can help  the transmitter, provided that the helper and the legitimate receiver have a channel advantage over the eavesdropper in the sense that $\frac{h_2}{\sigma_Y^2}> \frac{g_2}{\sigma_Z^2}$. In Figures \ref{case1_ervspe} and~\ref{case1_ervslk}, we evaluate the performance of our code design, and demonstrate this benefit of cooperation among the transmitter and the helper at finite blocklength. 

\textbf{Case 2}: $\frac{h_1}{\sigma_Y^2}> \frac{g_1}{\sigma_Z^2}$.  In this case, the legitimate receiver has a channel advantage over the eavesdropper. In Figures \ref{case1_ervspe} and \ref{case2_lk}, we evaluate the performance of our code design and, similar to Case 1, demonstrate that the helper can help the transmitter decreasing the information leakage at the eavesdropper, provided that the helper and the legitimate receiver have a channel advantage over the eavesdropper in the sense that $\frac{h_2}{\sigma_Y^2}> \frac{g_2}{\sigma_Z^2}$. 

Note that if $\frac{h_2}{\sigma_Y^2}\leq \frac{g_2}{\sigma_Z^2}$, then the helper could also help to improve the information leakage for the transmitter but would also negatively affect the probability of error of the secret message.

\subsubsection{Average probability of error}
\paragraph{With helper} \label{hs}

 \textit{Training}: We consider the channel model~\eqref{mod1a} {\color{black} with~$h_1=1$, $h_2=1$, and $\sigma^2_Y=1$.} For the design of the reliability layer described in Section \ref{rellyr}, the autoencoder is trained for $(n, q_1, q_2)=(12, 4, 4)$, $(n, q_1,q_2)=(16, 6, 6)$, $(n, q_1,q_2)=(20, 8, 8)$, and $(n, q_1, q_2)=(24,10, 10)$ over $600$ epochs with $2\cdot 10^5$ random input messages at learning rate $l_r=0.0001$ with a  batch size $b_s=500$.
 
 \textit{Testing}:
 To evaluate the average probability of error for the unprotected message $\mathbf{P}_e^{(M)}$, we first generate the inputs $V\in\{0,1\}^{q_1}$ and $M\in\{0,1\}^{q_2}$. Then, $V$ is passed through the trained encoder $e^0_{1,n}$ and $M$ is passed through the trained encoder $e^0_{2,n}$, which generates the codewords $X_1^n$ and $X_2^n$, respectively, and the channel output $Y^n$. Finally, the trained decoder $d^0_{2,n}$ forms an estimate of~$M$ from $Y^n$, as described in Section \ref{rellyr}. 

Consider $\varphi$ and $\psi$ with $q_1\in \{4,6,8,10\}$ and $n\in\{12,16,20,24\}$.
We chose the seeds as $\lambda\in \{0001, 000001, 00000001, 0000000001\}$ for the different values of $q_1$ and set the secret length $k_1 = 1$.  To evaluate the average probability of error for the secret message $\mathbf{P}_e^{(S)}$, the trained encoder $e^0_{1,n}$ encodes the message $S \in \{0,1\}^{k_1}$ as $e_{1,n}^0(\varphi(S,B))$, where $B \in \{0,1\}^{q_1-k_1}$ is a sequence of $q_1-k_1$ bits generated uniformly at random. The trained decoder $d^0_{1,n}$ forms $\widehat{S} \triangleq \psi(d^0_{1,n}(Y^n-\sqrt{h_2}\widehat{X}_2^n))$, as described in Section~\ref{SL}.

\paragraph{Without helper} \label{whs}
This scenario corresponds to $P_2=0$ and $M= \emptyset$  in the setting with a helper and can be implemented with point-to-point codes. Hence, to evaluate  $\mathbf{P}^{(S)}_e$, we used the best known point-to-point code construction for the Gaussian wiretap channel, i.e., the code design from \cite{rana2023short}.

\begin{figure}

        \centering

        \subfloat[]{

\includegraphics[trim=1.2cm 6.5cm 2.5cm 7cm,clip,width=0.45\textwidth]{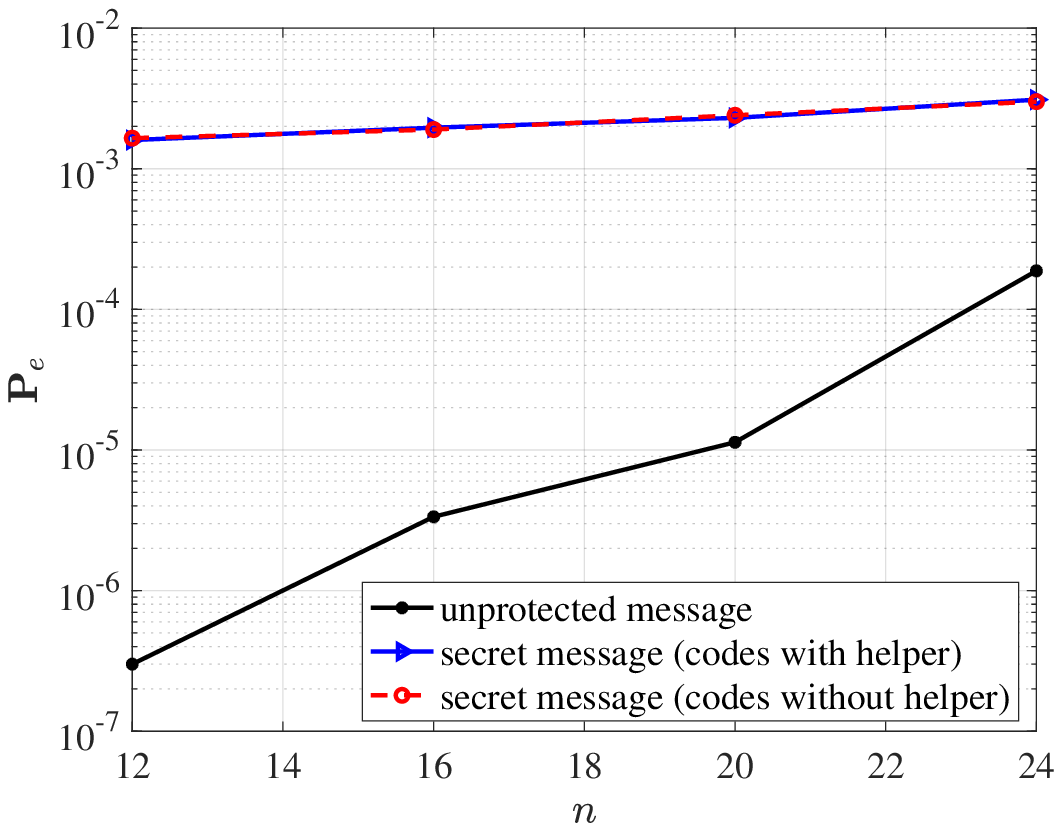}

        \label{case1_ervspe}

        }

       \centering

        \subfloat[]{

    \includegraphics[width=0.45\textwidth]{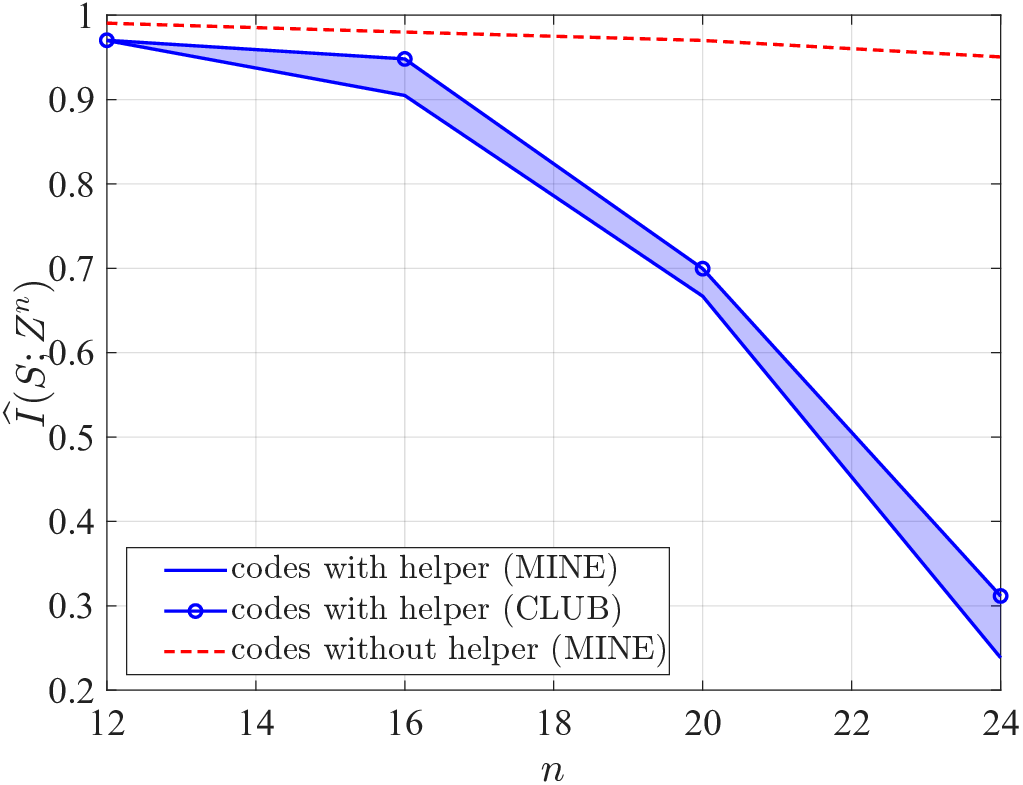}
    
\label{case1_ervslk}

        }

   \centering

    \subfloat[]{

    \includegraphics[width=0.45\textwidth]{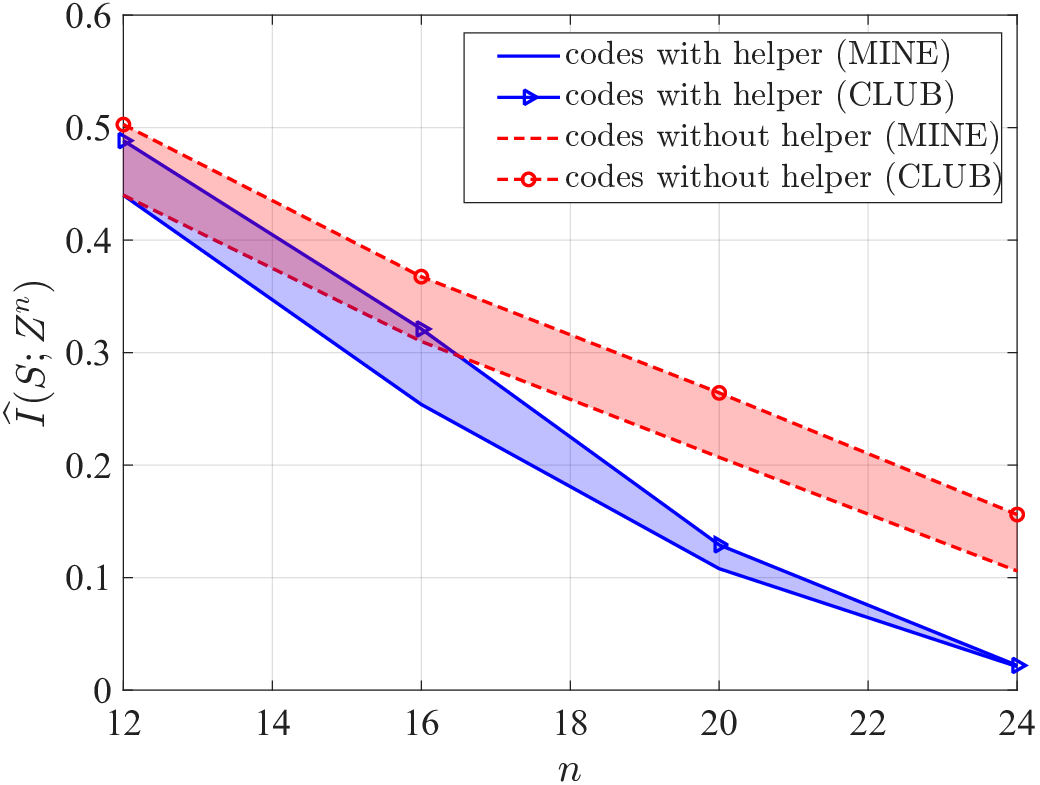}

         \label{case2_lk}}  

    \vspace{0cm}

        \caption{When~$\sigma^2_Y=1$, $\sigma^2_Z=1$, $h_1=1$, $h_2=1$, $(\frac{q_1}{n},\frac{q_2}{n})~\in~\{(\frac{4}{12},\frac{4}{12}), (\frac{6}{16},\frac{6}{16}), (\frac{8}{20},\frac{8}{20}), (\frac{10}{24},\frac{10}{24})\}$, 
        $k_1=1$,  $P_1=2$, and $P_2=12$. (a)~Average probability of error versus blocklength. (b)~Case~1: Information leakage versus blocklength obtained for $g_1=1$ and $g_2=0.3$. (c) Case 2: Information leakage versus blocklength obtained for $g_1=0.2$ and $g_2=0.3$.
        }

        \vspace{0cm}    

    \label{fig1}

    \end{figure}

Figure \ref{case1_ervspe} shows the average probability of error versus blocklength $n$. 
We observe a similar probability of error $\mathbf{P}^{(S)}_e$ for the secret message for both code designs with or without a helper, which shows that the presence of the helper does not degrade performance in terms of probability of error. Figure~\ref{case1_ervspe} also shows the probability of error $\mathbf{P}^{(M)}_e$ for the unprotected message, which is only present in code designs with a helper. 


\textcolor{black}{Note that  in all our simulations, we design the reliability layer with SIC, assuming $h_2P_2 \geq h_1P_1$.}

\subsubsection{Information leakage at the eavesdropper via MINE}
\label{mine}

\paragraph{With helper}\label{Lk_case1_fd}
We consider the model in \eqref{mod1b} with~$\sigma^2_Z~=~1$.  Set the secret length $k_1 = 1$, and set the unprotected message length $q_2\in\{4,6,8,10\}$.  Consider $\varphi$ and $\psi$ with $q_1\in \{4,6,8,10\}$ and $n\in\{12,16,20,24\}$.   We chose the seeds as $\lambda\in \{0001, 000001, 00000001, 0000000001\}$  for the different values of $q_1$.  Generate uniformly at random $S$ and $B$ that are fed to the encoder $e^0_{1,n}$ and outputs $X_1^n$. Similarly, generate uniformly at random $M$ that is fed to the encoder $e^0_{2,n}$ and outputs $X_2^n$. The output of encoders produces the channel outputs $Z^n$ at the eavesdropper.  To evaluate our lower bound of the leakage $I(S;Z^n)$, we use the MINE from \cite{r3a}. 
In particular, we use a fully connected feed-forward neural network with $4$ hidden layers, each having $400$ neurons and ReLU as an activation function. The input layer has $k_1+n$ neurons, and the ADAM optimizer with a learning rate of $0.0001$ is used for the training. We train the neural network over $100,000$ epochs of $20,000$ messages with a batch size of $2500$.
The samples of  the marginal distributions are obtained by either dropping $S$ or $Z^n$  from the joint samples $(S, Z^n)$ or shuffling the samples from the joint distribution along the batch axis. 

\paragraph{Without helper}
 To evaluate the leakage $I(S;Z^{n})$ for point-to-point codes, we used the code design from \cite{rana2023short}.

Figures \ref{case1_ervslk} and \ref{case2_lk} show the estimated information leakage versus blocklength~$n$ for Cases 1 and 2, respectively. 

\subsubsection{Information leakage at the eavesdropper via CLUB}
\label{club}
{\textcolor{black}
{With sample pairs $\{(s(i), z^n(i))\}_{i=1}^l$, $I(S;Z^n)$ has an unbiased estimation as
\begin{align}
    \widehat{I}_{\textup{CLUB}}& \triangleq \frac{1}{l}\textstyle\sum_{i=1}^{l} \log p(z^n(i)\vert s(i))\nonumber\\
    &\phantom{-----}-\frac{1}{l^2}\textstyle\sum_{i=1}^{l} \sum_{j=1}^{l} \log p(z^n(j)\vert s(i)),\label{clubeq}
\end{align}
where $\log p(z^n(i)\vert s(i))$ provides conditional log-likelihood of positive sample pair $(s(i), z^n(i))$ and $\log p(z^n(j)\vert s(i))$ provides conditional log-likelihood of negative sample pair $\{(s(i), z^n(j))\}_{i\neq j}$. To evaluate our upper bound of the leakage $I(S;Z^n)$, we use the CLUB estimator that
models the conditional distribution $p(z^n\vert s)=\mathcal{N}(Z^n; \mu(s),\sigma^2(s))$ \cite[Section 3.1]{cheng2020clubcontrastivelogratioupper}. 
In particular, we have two sub-networks, where one predicts $\mu(s)$ and the other predicts $\log \sigma^2(s)$, with each network having $3$ hidden layers, each having $400$ neurons and ReLU as an activation function. The input layer has $k_1$ neurons, the output layer has $n$ neurons, and the ADAM optimizer with a learning rate of $0.0008$ is used for the training. We train the neural network over $500,000$ epochs of $4000$ messages with a batch size of $2000$.}

{\textcolor{black}{Using positive sample pair $(s(i), z^n(i))$, we compute the first term in \eqref{clubeq} as $-\frac{1}{2l}\sum_{i=1}^{l}[ \frac{(z^n(i)-\mu(s(i)))^2}{\sigma^2(s(i))}+\log \sigma^2(s(i))+\log(2\pi)]$ and using negative sample pair $\{(s(i), z^n(j))\}_{i\neq j}$, we compute the second term in \eqref{clubeq} as $-\frac{1}{2l^2}\sum_{i=1}^{l}\sum_{j=1}^{l} [\frac{(z^n(j)-\mu(s(i)))^2}{\sigma^2(s(i))}+\log \sigma^2(s(i))+\log(2\pi)]$. }

{\textcolor{black}{Note that, for Case 1, without a helper, we did not include upper bounds because we want to show the gain obtain with the helpers in the worst case, i.e., we  considered the best leakage (lower bound) for the no-helper case.}

\subsubsection{Discussion}
As seen in Figures  \ref{case1_ervslk} and  \ref{case2_lk}, there is a significant improvement in terms of information leakage for codes with a helper compared to codes without a helper. Our proposed coding scheme with a helper helps to significantly improve information leakage when the transmitter has adverse channel conditions (Case 1), which is impossible to achieve with point-to-point codes. 
From Figures \ref{case1_ervspe} and  \ref{case1_ervslk},  we have designed codes (corresponding to Case 1) that show that the rate pair $(\frac{1}{24}, \frac{10}{24})$ is $(\epsilon_S=5.5\cdot 10^{-3},\epsilon_M= 4.2\cdot10^{-4},\textcolor{black}{\delta= 3.1\cdot 10^{-1})}$- achievable with power constraint $(2,12)$. Also, from Figures~\ref{case1_ervspe} and  \ref{case2_lk}, we designed codes (corresponding to Case 2) that show that the rate pair $(\frac{1}{24}, \frac{10}{24})$ is $(\epsilon_S=5.5\cdot 10^{-3}, \epsilon_M=4.2\cdot10^{-4}, \textcolor{black}{\delta=2.2\cdot 10^{-2}})$- achievable with power constraint $(2,12)$. Thanks to our modular approach, we only need to redesign the \textcolor{black}{security layer} for Case~1 and Case 2 since, in both cases, the channel gains $h_1$ and $h_2$ of the legitimate receiver's channel are unchanged.

Our codes' performance in Case 1 and Case 2 at short blocklengths are consistent with the previous work that considered non-constructive coding schemes and the asymptotic regime, e.g., \cite{tekin}, where a helper can improve the secrecy rate of a transmitter when either   $\frac{h_1}{\sigma_Y^2}> \frac{g_1}{\sigma_Z^2}$ or 
$\frac{h_1}{\sigma_Y^2}\leq \frac{g_1}{\sigma_Z^2}$.
{\color{black}
\subsection{Simulations and examples of code designs for $n \leq 64$}
We consider the channel model~\eqref{mod1a} {\color{black} with~$h_1=1$, $h_2=1$, and $\sigma^2_Y=1$.} For blocklengths $n\leq 64$, the autoencoder is trained for $(n, q_1, q_2)=(32, 8, 8)$, $(n, q_1,q_2)=(48, 10, 10)$, and  $(n, q_1,q_2)=(64, 11, 11)$ over $100$ epochs with $2\cdot 10^5$ random input messages at learning rate $l_r=0.0005$ with a  batch size $b_s=500$. Set the secret length $k_1 = 1$, and set the unprotected message length $q_2\in\{8,10,11\}$.  Consider $\varphi$ and $\psi$ with $q_1\in \{8,10,11\}$ and $n\in\{32,48,64\}$.   We chose the seeds as $\lambda\in \{00000001, 0000000001, 00000000001\}$  for the different values of $q_1$. Then, $400\times 10^5$ random messages are used to evaluate the average probability of errors $\mathbf{P}_e^{(M)}$ and $\mathbf{P}_e^{(S)}$ as described in Section \ref{hs}.  When there is no helper, the scenario corresponds to $P_2=0$ and $M= \emptyset$  in the setting with a helper as described in Section \ref{whs}. Figure \ref{fig:lb_er} shows the average probability of error versus blocklength $n$. 

     \begin{figure}[ht]
    \centering
     \includegraphics[width=1\linewidth]{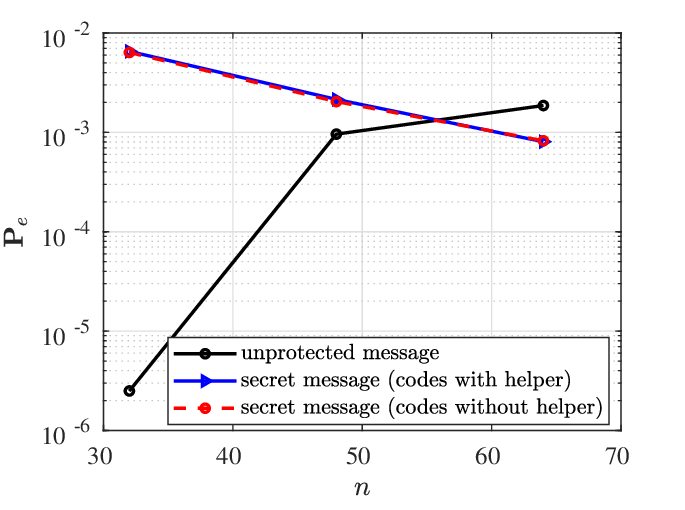}
     \caption{Average probability of error versus blocklength. When~$\sigma^2_Y=1$,  $h_1=1$, $h_2=1$, $(\frac{q_1}{n},\frac{q_2}{n})~\in~\{(\frac{8}{32},\frac{8}{32}), (\frac{10}{48},\frac{10}{48}), (\frac{11}{64},\frac{11}{64})\}$, 
        $k_1=1$,  $P_1=1$, and $P_2=6$. 
         }
     \label{fig:lb_er}
 \end{figure}

The \textcolor{black}{security layer} is implemented similar to Section \ref{SL}   with $k=1$, and we compute the leakage $I(S;Z^n)$ as in \textcolor{black}{Sections \ref{mine} and \ref{club}}. We consider the model in \eqref{mod1b} with~$\sigma^2_Z~=~1$.  Figure \ref{fig:lb_lk} shows the information leakage versus blocklength~$n$ for Case~1. 
     
 \begin{figure}[ht]
    \centering
     \includegraphics[width=.9\linewidth]{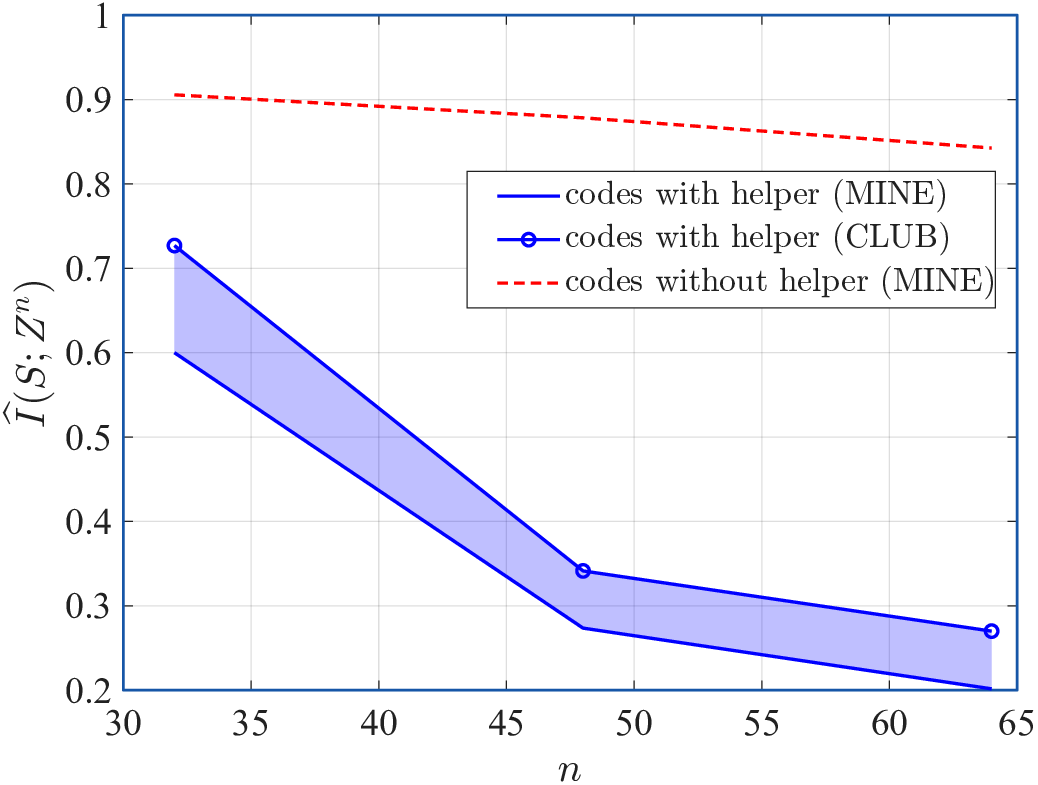}
     \caption{Information leakage versus blocklength obtained for $g_1=1$ and $g_2=0.9$. When~$\sigma^2_Y=1$, $\sigma^2_Z=1$, $h_1=1$, $h_2=1$, $(\frac{q_1}{n},\frac{q_2}{n})~\in~\{(\frac{8}{32},\frac{8}{32}), (\frac{10}{48},\frac{10}{48}), (\frac{11}{64},\frac{11}{64})\}$, 
         $k_1=1$,  $P_1=1$, and $P_2=6$. 
         }
     \label{fig:lb_lk}
 \end{figure}

}
\section{Multiple-helper-Assisted Code Design based on point-to-point codes}\label{gwtc_arch2}

In Section \ref{rellyr}, in the design of the reliability layer,  the receiver successively cancels the interference in order to estimate the messages. However, this setting becomes complex as the number of transmitters increases in the network because the receiver needs to subtract each interference caused by transmitters, which requires additional calculations. To reduce  complexity, we propose a simplified training method that does not rely on SIC but rather on multiple independent decoders. During testing, however, the receiver employs a SIC architecture to decode the messages. 

Next, we describe the model in Section \ref{model_helpers_GWT}. We present our
coding scheme design in Section \ref{rellyr_arch2} and we evaluate the
performance of our code design through simulations for the
Gaussian wiretap channel with multiple helpers
in Section~\ref{sim_helpers}. 

\subsection{Model}\label{model_helpers_GWT}
Consider a Gaussian wiretap channel with one transmitter and  $L-1$ helpers indexed in  $\mathcal{J}\triangleq\{2,\dots, L\}$, which  is defined~by
\begin{align}
    Y&\triangleq\sqrt{h_1}X_1+\sum_{j=2}^{L}\sqrt{h_j}X_j+N_Y\label{mod1aarc2},\\
    Z&\triangleq\sqrt{g_1}X_1+\sum_{j=2}^{L}\sqrt{g_j}X_j+N_Z\label{mod1barc2},
\end{align}
where $h_1$ and $h_j$ are the channel gains of the transmitter and $j$-th helper, respectively, to the intended receiver, $g_1$ and $g_j$ are the channel gains of the transmitter and $j$-th helper, respectively, to the eavesdropper, and $N_Y$ and $N_Z$ are zero-mean Gaussian random variables with variances $\sigma^2_Y$ and $\sigma^2_Z$, respectively. The objective of the transmitter is to send a secret message to the receiver with a low probability of error by cooperating with helpers.  Additionally, each helper wishes to transmit an unprotected message to the receiver with a low probability of error.
\subsection{Design of reliability layer}\label{rellyr_arch2}
The reliability layer for $L$ users consists of designing a code described by $L$ encoders and $L$ decoders for the channel described in \eqref{mod1aarc2}.  The design of the reliability layer is based on neural network autoencoders{ \color{black}that are trained using Algorithm~\ref{alg2}.} Define $\mathcal{L} \triangleq \{1, 2, \dots, L\}$.
As depicted in Figure~\ref{MAC_S2_M2}, the input message $V_1\in \mathcal{V}_1\triangleq \{1, 2, \dots, 2^{q_1}\}$  is encoded using a neural network encoder $e^{0}_{1,n}$ to obtain the codeword $X_1^n$, and the input messages $(M_j\in\mathcal{M}_j\triangleq \{1, 2, \dots, 2^{q_j}\})_{j\in\mathcal{J}}$ are encoded using  neural network encoders $(e^0_{j,n})_{j\in\mathcal{J}}$ to obtain the codewords $(X^n_j)_{j\in\mathcal{J}}$. During training, the decoder $d^{0}_{j,n}$ receives 
\begin{align*}
    Y_j^n\triangleq \sum_{j'=1}^{j}\sqrt{h_{j'}}X^n_{j'}+N_Y^n,~j\in\mathcal{J},
\end{align*}
   where $N_Y\sim\mathcal{N}(0,\sigma^2_Y)$ and recovers $M_j$  as $\widehat{M}_j$, while treating 
$\sum_{j'=1}^{j-1}\sqrt{h_{j'}}X^n_{j'}$ 
as noise, as written in Line 4. 
Then, as depicted in Figure~\ref{MAC_S3_M2}, the input message $V_1$ is encoded using a neural network encoder $e^{0}_{1,n}$ to obtain the codeword $X_1^n$, and  from the channel outputs  $Y_1^n\triangleq X^n_1+N^n_Y$, the decoder $d^0_{1,n}$ recovers $V_1$ as $\widehat{V}_1$, as written in Line 6. {\color{black} Without loss of generality one can relabel the users such that $h_LP_L\geq h_{L-1}P_{L-1}\geq\dotsc\geq h_1P_1$.} The autoencoder is trained over all possible messages using an ADAM optimizer, and the individual categorical cross-entropy loss function is denoted by $\text{loss}_j\triangleq-\log(p_{m_j})$, ${j\in\mathcal{J}}$, and $\text{loss}_1\triangleq-\log(p_{v_1})$ of the helper-receiver and the transmitter-receiver pair, respectively, where $m_j\in \mathcal{M}_j$,   and $v_1\in\mathcal{V}_1$, as written in Line 10.   This training algorithm solely relies on point-to-point codes. Note that for testing, as described in Section \ref{sim_helpers}, the decoder employs SIC.

\begin{algorithm}
	\caption{Proposed training algorithm in Section \ref{rellyr_arch2} based on point-to-point codes }
 \textbf{Require} Number of epochs $N_e$, Batch size $b_s$, Learning rate~$l_r$
\begin{algorithmic}[1] 
\State Define  the autoencoder model for Helper  $j\in\mathcal{J}$, with encoder
$e^0_{j,n}$,
and decoder $d^0_{j,n}$, where the input to the decoder  is 
$$Y_j^n\triangleq\sum_{j'=1}^{j}\sqrt{h_{j'}}X^n_{j'}+N_Y^n, $$  where ${X}^n_{j'}\triangleq e_{j',n}^0({M}_{j'})$, as shown in  Figure~\ref{MAC_S2_M2}. 
\State Define the autoencoder model for the transmitter as shown in  Figure~\ref{MAC_S3_M2} with an encoder $e^0_{1,n}$ and decoder $d^0_{1,n}$, where the input to the decoder $d^0_{1,n}$ is $$Y_1^n\triangleq X_1^n+N_Y^n.$$ The hyperparameters of the autoencoder are described in Section~\ref{rellyr}.
\For {$j=L~\text{to}~2$}
\State 
Use decoder $d^0_{j,n}$ to estimate message $M_j$ from $Y_j^n$.
\EndFor
\State 
Use decoder $d^0_{1,n}$ to estimate message $V_1$ from $Y_1^n$.
    \For{$i \leq N_e$}
\State Generate random input messages $\{V_1,M_2,\dots, M_L\}$ and generate noise $N_Y\sim \mathcal{N}(0,\sigma_Y^2)$.
\State Train the $L$ autoencoder models using  inputs $\{V_1,M_2,\dots, M_L\}$.
\State Individual loss   $\text{loss}_l$ is  minimized for any  $l\in\mathcal{L}$.
\EndFor

	\end{algorithmic} \label{alg2}
\end{algorithm}

\begin{figure}[ht]
\centering
\includegraphics[trim=2cm 1cm 2cm 1cm,clip, width=0.5\textwidth]{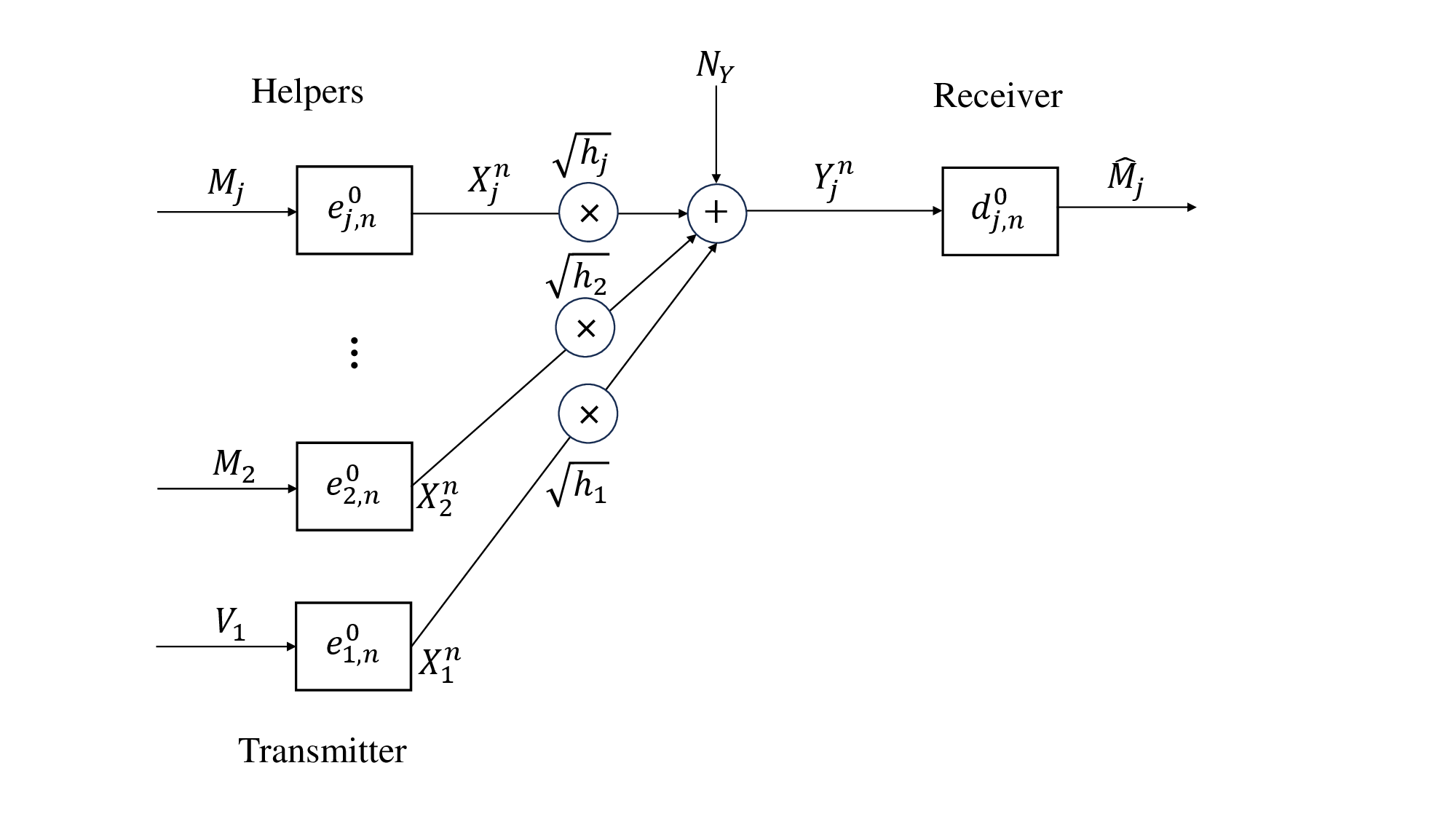}

\caption{Training architecture of the autoencoder for any helper  $j\in\mathcal{J}$ based on point-to-point codes.} \label{MAC_S2_M2}
\end{figure}

\begin{figure}[ht]
\centering
\includegraphics[trim=4cm 6cm 4cm 5cm,clip, width=0.5\textwidth]{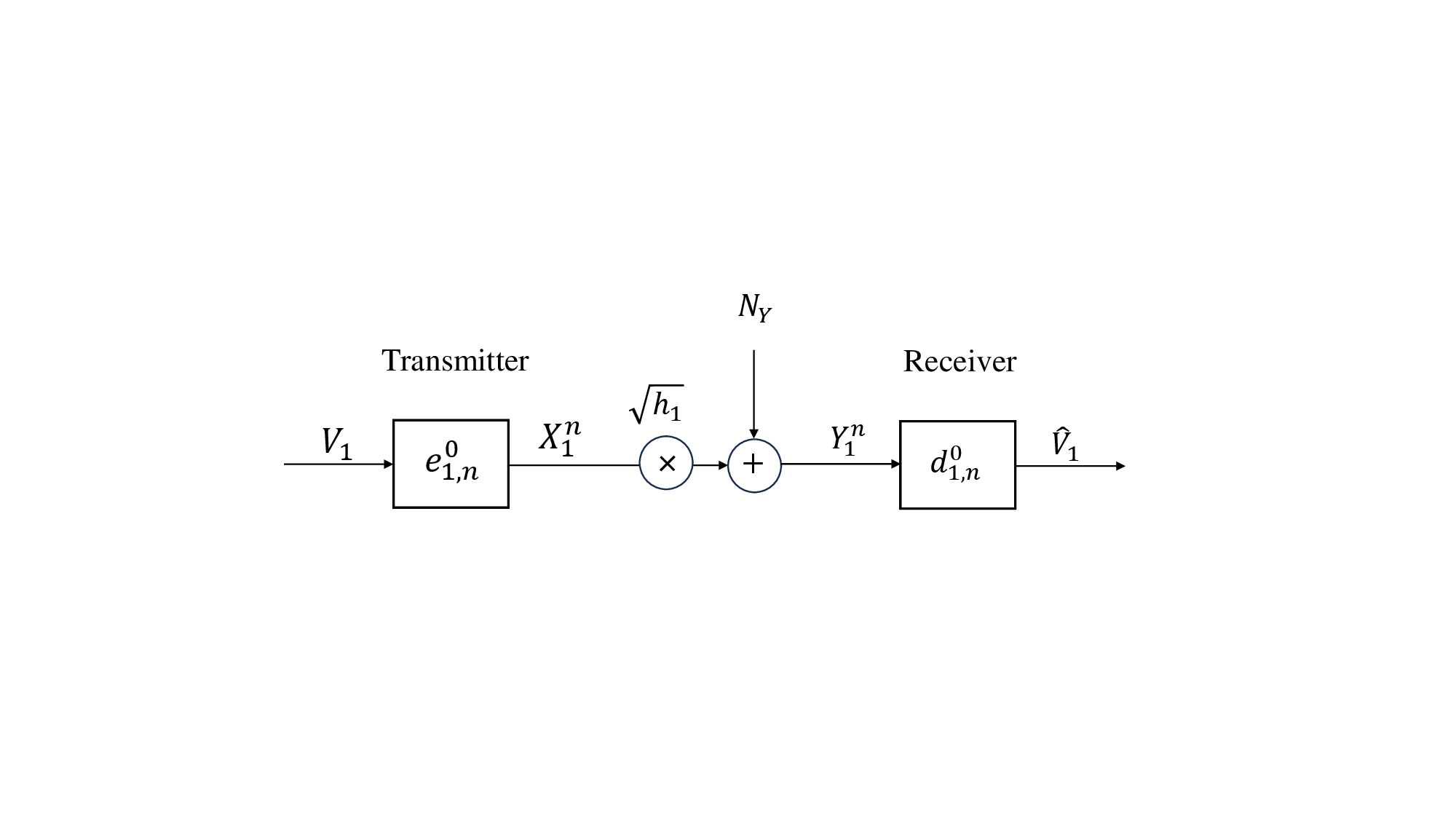}

\caption{Training architecture of the autoencoder for the transmitter based on point-to-point codes.} \label{MAC_S3_M2}
\end{figure}

The generalized training algorithm for $L$ users (one transmitter and $L-1$ helpers) indexed in $\mathcal{L}$ based on Section~\ref{rellyr} is shown in Algorithm~\ref{MAC_alg1}. The main difference between Algorithm~\ref{alg2} and Algorithm~\ref{MAC_alg1} is the decoder architecture, which results in the former minimizing individual loss between each input and the estimated output. 

\begin{algorithm}
	\caption{Proposed    training algorithm based on Section~\ref{rellyr}}
\textbf{Require} Number of epochs $N_e$, Batch size $b_s$, Learning rate~$l_r$
\begin{algorithmic}[1]
\State Define the autoencoder model with encoders $(e^0_{l,n})_{l\in\mathcal{L}}$
and decoders $(d^0_{l,n})_{l\in\mathcal{L}}$, where the input to the decoder $d^0_{l, n}$, $l\in \mathcal{L}$, is  
$$Y_l^n\triangleq Y^n-\sum_{j'=l+1}^{L}\sqrt{h_{j'}}\widehat{X}^n_{j'}, $$ 
 where $\widehat{X}^n_{j'}\triangleq e_{j',n}^0(\widehat{M}_{j'})$, as shown in Figure \ref{MAC_jo}. The hyperparameters of the autoencoder are described in Section~\ref{rellyr}.
\For {$l=L~\text{to}~2$}
\State Use decoder $d_{l,n}$ to estimate message $M_l$ from $Y_l^n$.
\EndFor	
\State Use decoder $d_{1,n}$ to estimate message $V_1$ from $Y_1^n$.
     \For {$i \leq N_e$}
\State Generate random input messages $\{V_1,M_2,\dots, M_L\}$ and generate noise $N_Y\sim \mathcal{N}(0,\sigma_Y^2)$.
\State Train the autoencoder model using inputs $\{V_1,M_2,\dots, M_L\}$ .
\State The sum of all individual losses, i.e., $\sum_{l=1}^{L} \text{loss}_l$,  is minimized. 
\EndFor
 
	\end{algorithmic} \label{MAC_alg1}
\end{algorithm}

\begin{figure}[ht]
\centering
\includegraphics[trim=2cm 0cm 2cm 0cm,clip, width=0.5\textwidth]{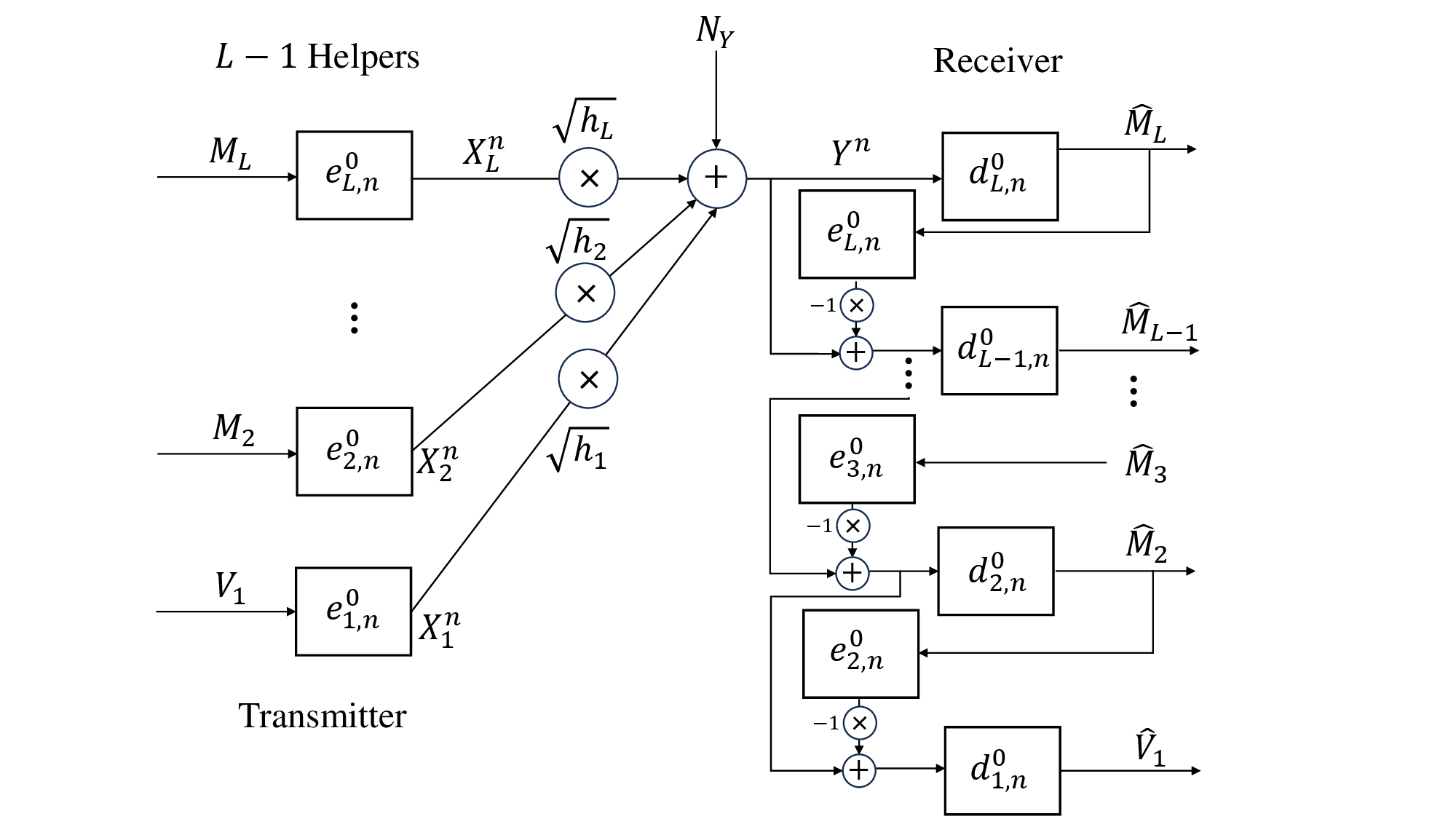}

\caption{ Training architecture of the autoencoder for $L$ users based on SIC.} \label{MAC_jo}
\end{figure}

The design of the \textcolor{black}{security layer} is similar to that of Section~\ref{SL}. 

\subsection{Simulations}\label{sim_helpers}
\subsubsection{Average probability of error}\label{sec:error_arch2}
We consider the channel model \eqref{mod1aarc2} {\color{black} with~$\sigma^2_Y=1$, $h_1=1$, $h_j=1$ for  $j\in \mathcal{J}$ and $L\in\{1,2,3,4\}$}. Set the helper  message length $q_j\in\{4,6,8,10\}$, where $j\in  \mathcal{J}$, and $n\in\{12,16,20,24\}$. For the design of the reliability layer as described in Algorithm~\ref{alg2}, for $j\in \mathcal{J}$,  the autoencoders are trained for $(n, q_1, q_j)=(12,4, 4)$, $(n, q_1, q_j)=(16, 6,6)$, $(n, 
 q_1, q_j)=(20, 8, 8)$, and $(n, q_1, q_j)=(24,10, 10)$  over $600$ epochs with $2\cdot 10^5$ random input messages at $l_r=0.0001$ with   $b_s=500$. Note that when $L=1$, there is no helper, and it can be implemented with point-to-point codes, as discussed in Section \ref{whs}.
\begin{figure}[ht]
    \centering
    \includegraphics[trim=1.2cm 6.5cm 2.5cm 7cm,clip,width=0.45\textwidth]{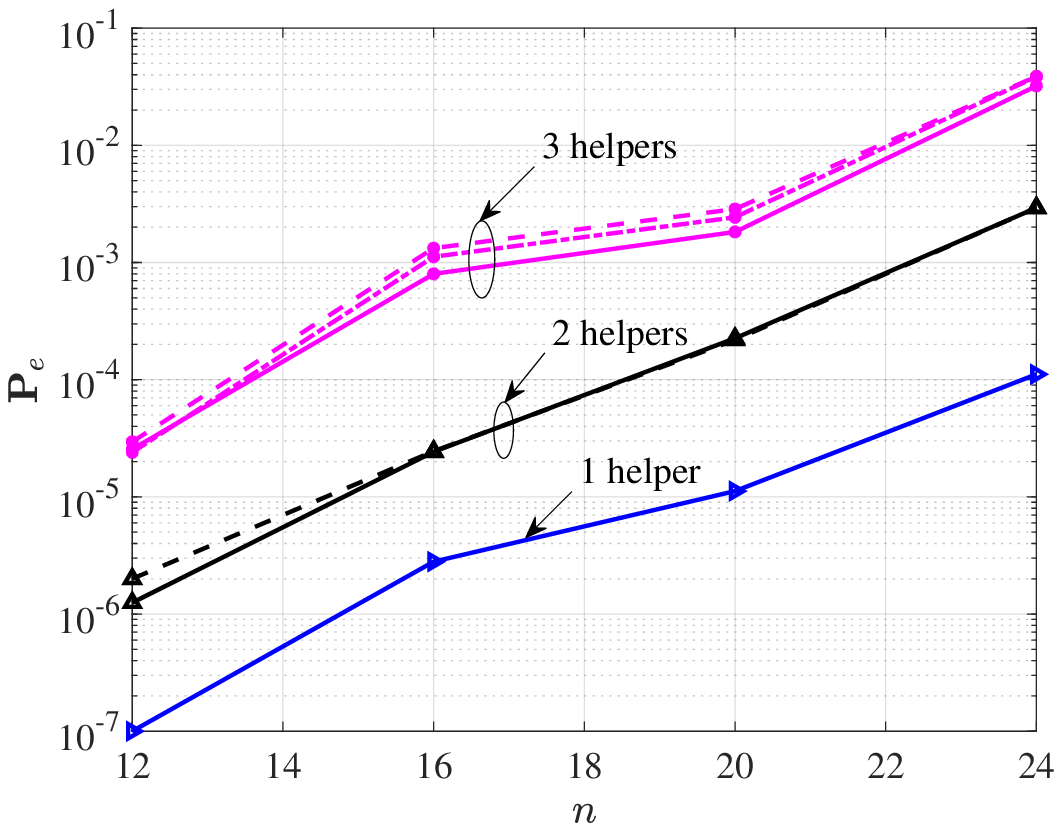}
    \caption{Average probability of error $\mathbf{P}_e^{(M_j)}$ versus blocklength. $\sigma^2_Y=1$, $\sigma^2_Z=1$, $P_1=2$, and ~$P_j=12$, $j\in\{2,3,4\}$. The channel gains of the transmitter are $h_1=1$ and $h_j=1$, $j\in\{2,3,4\}$.}
    \label{messageerror_multihelpers}
\end{figure}

\begin{figure}[ht]
    \centering
    \includegraphics[trim=1.2cm 6.5cm 2.5cm 7cm,clip,width=0.45\textwidth]{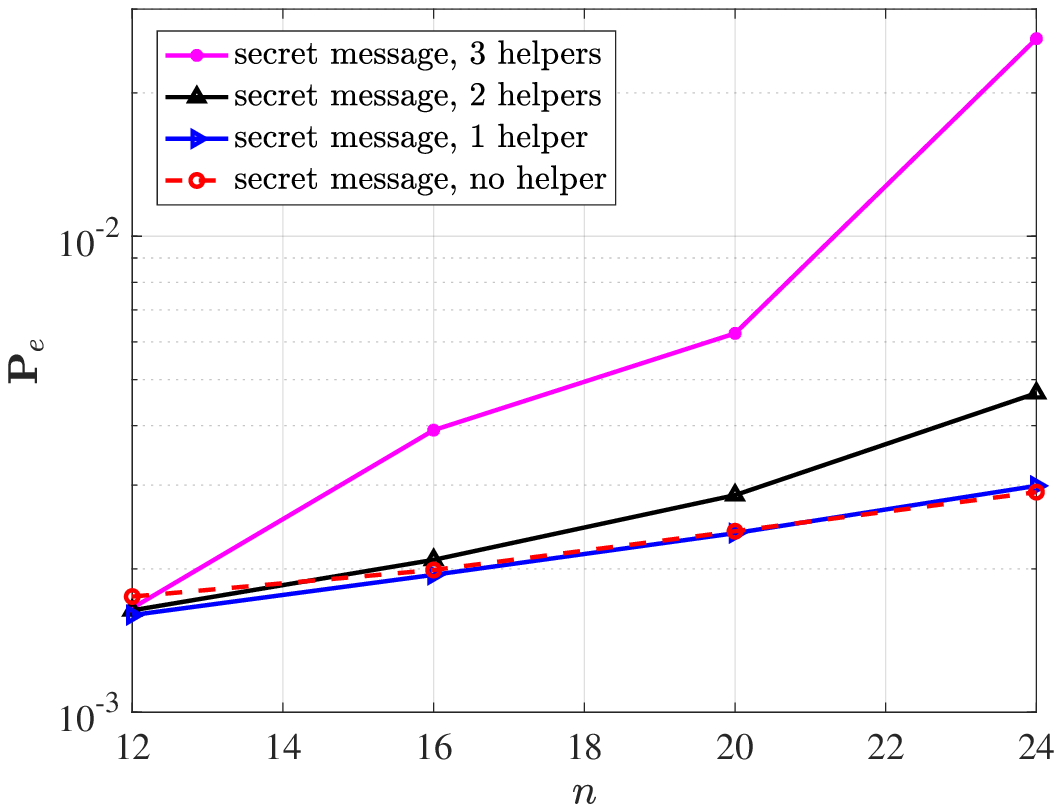}
    \caption{Average probability of error $\mathbf{P}_e^{(S)}$ versus blocklength when $k_1=1$. $\sigma^2_Y=1$, $P_1=2$, and ~$P_j=12$, $j\in\{2,3,4\}$. The channel gains of the transmitter are $h_1=1$ and $h_j=1$, $j\in\{2,3,4\}$.}
    \label{secreterror_multihelpers}
\end{figure}

\textit{Testing}:
 To evaluate the average probability of error $\mathbf{P}_e^{(M_j)}$, $j\in\mathcal{J}$, we first generate the inputs $V_1\in\{0,1\}^{q_1}$ and $M_j\in\{0,1\}^{q_j}$, $j\in\mathcal{J}$. Then, $V_1$ is passed through the trained encoder $e^0_{1,n}$ and $M_j$ is passed through the trained encoder $e^0_{j,n}$, which generates the codewords $X_1^n$ and $X_j^n$, respectively, and the channel output
 \begin{align*}
     Y^n\triangleq \sqrt{h_1}X^n_1+\sum_{j=2}^{L}\sqrt{h_j}X^n_j+N^n_Y.
 \end{align*}
  Finally, the trained decoder $d^0_{j,n}$ forms an estimate $\widehat{M}_{j}$ of~$M_j$,  $~\text{for}~j = L~\text{to}~2$ from 
 \begin{align*} 
     Y^n-\sum_{j'=j+1}^{L}\sqrt{h_{j'}}\widehat{X}^n_{j'},
 \end{align*}
 where $\widehat{X}^n_{j'}\triangleq e_{j',n}^0(\widehat{M}_{j'})$.

Consider $\varphi$ and $\psi$ with the transmitter message length $q_1\in \{4,6,8,10\}$ and $n\in\{12,16,20,24\}$.
We chose the seeds as $\lambda\in \{0001, 000001, 00000001, 0000000001\}$ for the different values of $q_1$ and set the secret length $k_1 = 1$. 
Further, to evaluate the average probability of error $\mathbf{P}_e^{(S)}$ for the secret message $S$, the trained encoder $e^0_{1,n}$ encodes the message $S \in \{0,1\}^{k_1}$ as $e_{1,n}^0(\varphi(S,B))$,  where $B \in \{0,1\}^{q_1-k_1}$ is a sequence of $q_1-k_1$ bits generated uniformly at random. The trained decoder $d^0_{1,n}$ forms 
\begin{align*}
    \widehat{S} \triangleq \psi(d^0_{1,n}(Y^n-\textstyle\sum_{j=2}^L\sqrt{h_j}\widehat{X}_j^n)),
\end{align*}
as described in Section~\ref{SL}. 
Figure~\ref{messageerror_multihelpers}  shows the probability of error $\mathbf{P}^{(M_j)}_e$, $j\in\mathcal{J}$,  for the unprotected messages for a different number of helpers. Figure \ref{secreterror_multihelpers} shows the average probability of error for the secret message $\mathbf{P}_e^{(S)}$. We observe that the probability of error increases as the number of helpers increases due to codewords transmitted by the helpers being treated as noise. 

\subsubsection{Information leakage at the eavesdropper}\label{sec:lk_arch2}
 For the simulations, we consider the model in \eqref{mod1barc2} with $\sigma_Z^2=1$ and secret length $k_1=1$. {\color{black}We have $L\in\{1,2,3,4\}$, $g_1=1$, $g_j=0.3$,  helper  message length $q_j\in\{4,6,8,10\}$, where $j\in \mathcal{J}$ and $n\in\{12,16,20,24\}$}. Consider $\varphi$ and $\psi$ with the transmitter message length $q_1\in\{4,6,8,10\}$. We chose the seeds as $\lambda\in \{0001, 000001, 00000001, 0000000001\}$ for the different values of $q_1$.  We compute the leakage $I(S;Z^{n})$ for the helper-assisted code design based on point-to-point codes, similar to \textcolor{black}{Sections \ref{mine} and \ref{club}} in Figure~\ref{fig:leakage_multihelpers_arc2} for different number of helpers. We observe that the information leakage decreases as the number of helpers increases. This is because the codeword transmitted by a helper can be seen as a jamming signal from the eavesdropper's point of view. An increased number of helpers equates to more jamming power, which results in lower information leakage.

\begin{figure}
    \centering
    \includegraphics[width=.45\textwidth]{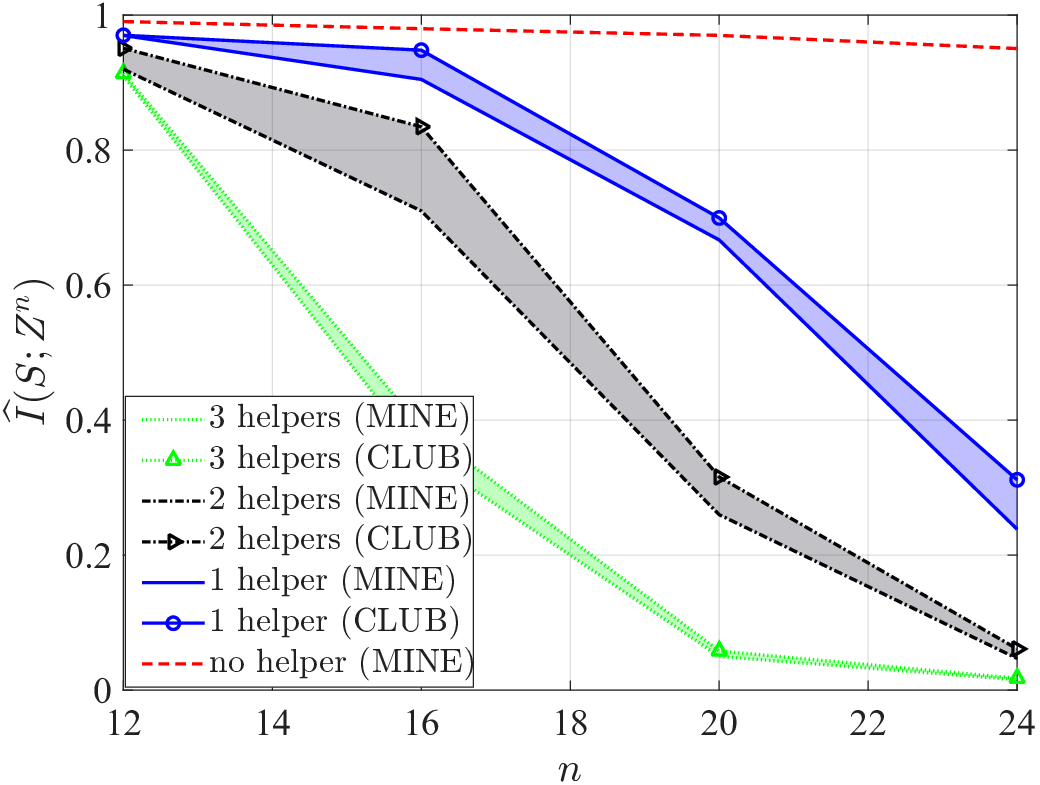}  
    \caption{Information leakage versus blocklength obtained for $g_1=1$ and $g_j=0.3$, $j\in\{2,3,4\}$ when $k_1=1$. $\sigma^2_Y=1$, $\sigma^2_Z=1$, $P_1=2$, and ~$P_j=12$, $j\in\{2,3,4\}$. 
    }
    \label{fig:leakage_multihelpers_arc2}
\end{figure}
{\color{black}In Figure~\ref{fig:error_leakage},  we plotted the average probability of error $\mathbf{P}_e^{(S)}$ versus information leakage  obtained from Figures~\ref{secreterror_multihelpers} and~\ref{fig:leakage_multihelpers_arc2}. This highlights that the
optimal number of helpers is determined by the total transmit power from the helpers
and the channel gains of the helpers at the intended receiver and eveasdropper.} 
\begin{figure}
    \centering   \includegraphics[width=0.45\textwidth]{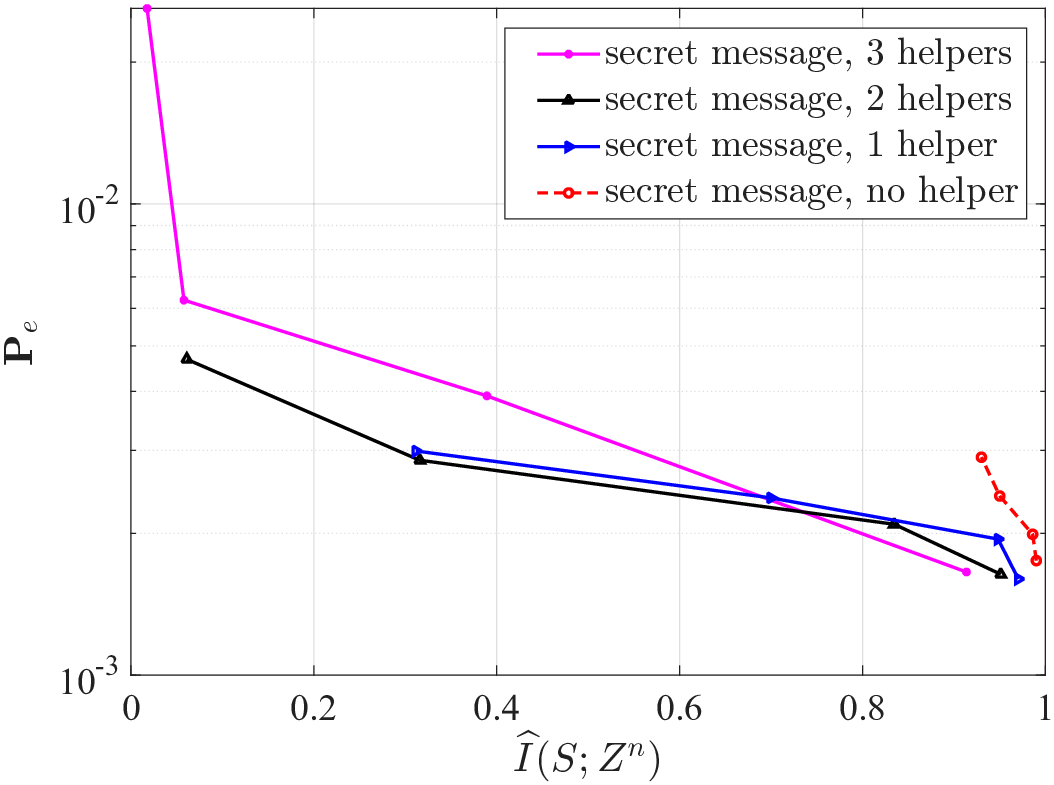}
    \caption{Average probability of error $\mathbf{P}_e^{(S)}$ versus information leakage. Average probability of error is obtained for $k_1=1$, $\sigma^2_Y=1$, $P_1=2$, and ~$P_j=12$, $j\in\{2,3,4\}$. The channel gains of the transmitter are $h_1=1$ and $h_j=1$, $j\in\{2,3,4\}$. Information leakage is obtained for $g_1=1$ and $g_j=0.3$, $j\in\{2,3,4\}$ when $k_1=1$. $\sigma^2_Y=1$, $\sigma^2_Z=1$, $P_1=2$, and ~$P_j=12$, $j\in\{2,3,4\}$. 
    }
 \label{fig:error_leakage}
\end{figure}

\subsubsection{Time complexity comparison between Algorithms~\ref{MAC_alg1} and~\ref{alg2}}

{\color{black} We now compare the helper-assisted code design based on point-to-point codes described in Algorithm~\ref{alg2} with the code design described in Algorithm~\ref{MAC_alg1}. 
Figure \ref{fig:timecomp} compares the training times of Algorithm \ref{alg2} and Algorithm \ref{MAC_alg1}, showing that Algorithm \ref{alg2} requires less time. Due to the simplified decoding architecture shown in Figure~\ref{MAC_S3_M2}, the code design based on point-to-point codes has a significant time gain compared to the code design outlined in Algorithm~\ref{MAC_alg1}. The time is measured while training autoencoders in the reliability layer for the parameters given in Section~\ref{sec:error_arch2}. For the same parameter, we observe a similar average probability of error for both code designs, as shown in Table \ref{onehelper}. }
\begin{table}[h!]
\begin{center}
\caption{Average probability of error $\mathbf{P}_e^{(S)}$ comparison between code designs.}
\begin{tabular}{ | m{3em} | m{1.8cm}| m{1.8cm} | } 
  \hline
  $(n,k_1)$& Algorithm~\ref{alg2} $\mathbf{P}_e^{(S)}$($\vert \mathcal{J}\vert=1$) & Algorithm~\ref{MAC_alg1} $\mathbf{P}_e^{(S)}$($\vert \mathcal{J}\vert=1$) \\ 
  \hline
  $(12,4)$ & $1.65\cdot 10^{-3}$ & $1.68\cdot 10^{-3}$\\ 
  \hline
  $(16,6)$ & $1.96\cdot 10^{-3}$ & $1.94\cdot 10^{-3}$ \\ 
  \hline
  $(20,8)$ & $2.33\cdot 10^{-3}$ & $2.37\cdot 10^{-3}$ \\ 
  \hline
  $(24,10)$ & $3.10\cdot 10^{-3}$ & $2.99\cdot 10^{-3}$ \\ 
  \hline
\end{tabular}\label{onehelper}
\end{center}
\end{table}

\begin{figure}[ht]
    \centering
    \includegraphics[trim=0cm 0cm 0cm 0cm,clip,width=0.45\textwidth]{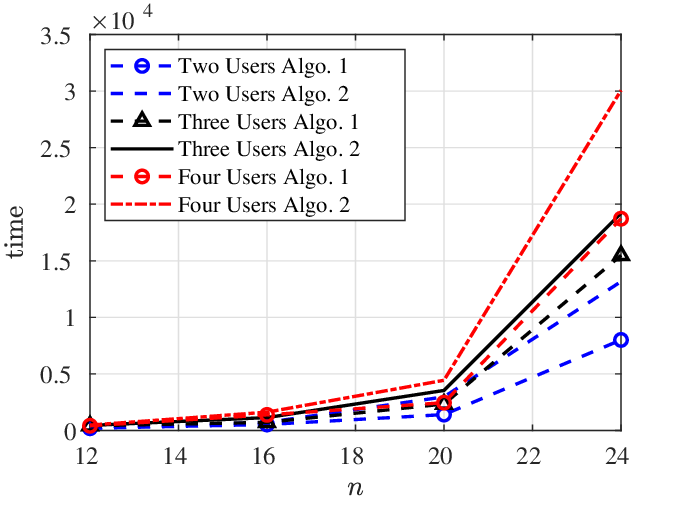}
    \caption{Comparison of training times (in seconds) between Algorithms~\ref{alg2} and  \ref{MAC_alg1}. $\sigma^2_Y=1$, $P_1=2$,~ and ~$P_j=12$, $j\in\{2,3,4\}$. The channel gains of the transmitter are $h_1=1$ and $h_j=1$, $j\in\{2,3,4\}$.}
    \label{fig:timecomp}
\end{figure}

From Figures \ref{messageerror_multihelpers}, \ref{secreterror_multihelpers}, and \ref{fig:leakage_multihelpers_arc2}, we designed codes that show that $\Big(\frac{1}{24}, \frac{10}{24}, \frac{10}{24}\Big)$ is $(\epsilon_S= 4.9\cdot 10^{-3},\epsilon_{M_2}=2.9\cdot 10^{-3}, \epsilon_{M_3}=2.7\cdot 10^{-3},\textcolor{black}{ \delta=6.1 \cdot 10^{-2} })$- achievable with power constraints $(2,12,12)$.

\section{Helper-Assisted Code Design when two transmitters send secret messages}\label{mac_h}

In Sections \ref{model_des}-\ref{gwtc_arch2},  only one transmitter communicates a confidential message to the receiver in the presence of an eavesdropper.  In this section, we consider the Gaussian multiple access wiretap channel with multiple transmitters and~helpers.

 We first introduce the Gaussian multiple access wiretap channel with helpers in Section~\ref{mac_h_mod}. We then describe our code design and evaluate its performance through simulations in Sections \ref{mac_h_cd} and \ref{mac_h_sim}, respectively.

\subsection{Model}\label{mac_h_mod}
  A  Gaussian multiple access wiretap channel with two transmitters and $L-2$ helpers is defined by
\begin{align}
    Y&\triangleq\sum_{i=1}^{2}\sqrt{h_i}X_i+\sum_{j=3}^{L}\sqrt{h_j}X_j+N_Y\label{mod1aarc3},\\
    Z&\triangleq\sum_{i=1}^{2}\sqrt{g_i}X_i+\sum_{j=3}^{L}\sqrt{g_j}X_j+N_Z\label{mod1barc3},
\end{align}
where $h_i$, and $h_j$ are the channel gains of the $i$-th transmitter and $j$-th helper, respectively, to the intended receiver, $g_i$ and $g_j$ are the channel gains of the $i$-th transmitter and $j$-th helper, respectively, to the eavesdropper, and $N_Y$ and $N_Z$ are zero-mean Gaussian random variables with variances $\sigma^2_Y$ and $\sigma^2_Z$, respectively. The objective of both transmitters is to each send a confidential message to the intended receiver with a low probability of error by cooperating with the helpers. Additionally, each helper also wishes to send an unprotected message to the receiver with a low error probability.   
\begin{definition}
 Define $\mathcal{I}\triangleq \{1,2\}$ and $\mathcal{J}\triangleq\{3,\dots,L\}$. A $((k_l)_{l\in \mathcal{L}}, n, (P_l)_{l\in \mathcal{L}})$ Gaussian multiple access wiretap channel code with helpers consists of
\begin{itemize}
    \item $L$ encoders:
    \begin{itemize}
        \item an encoder for the $i$-th transmitter, $i\in\mathcal{I}$, \\
        $e_{i,n}:\{0,1\}^{k_i}\rightarrow \mathbb{B}^n_0(\sqrt{nP_i})$, which, for a message $S_i\in \{0,1\}^{k_i}$, forms the codeword~$X_i^n\triangleq~e_{i,n}(S_i)$;
          \item an encoder for the $j$-th helper,   $j\in\mathcal{J}$, \\
          $e_{j,n}:\{0,1\}^{k_j}\rightarrow \mathbb{B}^n_0(\sqrt{nP_j})$, which, for a message $M_j\in \{0,1\}^{k_j}$, forms the codeword~$X_j^n\triangleq~e_{2,n}(M_j)$;
    \end{itemize}
    \item a decoder
        $d_{n}: \mathbb{R}^n\rightarrow \{0,1\}^{\sum^2_{i=1}k_i}\times  \{0,1\}^{\sum^L_{j=3}k_j}$,
        which, from the channel observations $Y^n$, forms an estimate of the transmitted messages $((S_i)_{i\in\mathcal{I}},(M_j)_{j\in\mathcal{J}})$ as $d_{n}(Y^n)$;
       \end{itemize}
The codomain of the encoders reflects the average power constraints given by 
\begin{align*}
    \sum_{t=1}^{n}(X_l(t))^2\leq nP_l,~l\in\mathcal{L}.
\end{align*}
\end{definition}

The performance of a $((k_l)_{l\in \mathcal{L}}, n, (P_l)_{l\in \mathcal{L}})$ code is measured in terms of
\begin{enumerate}
    \item The average probability of error for the secret message~$S_i$, $i\in\mathcal{I}$,
    \begin{align}
        \mathbf{P}^{(S_i)}_e\triangleq \frac{1}{2^{k_i}}\sum^{2^{k_i}}_{s_i=1}\mathbb{P}[ \widehat{S}_i\neq s_i\vert s_i~\text{is~sent} ]; 
    \end{align}
    \item The average probability of error for the unprotected message~$M_j$, $j\in\mathcal{J}$,
    \begin{align}
     \mathbf{P}^{(M_j)}_e\triangleq \frac{1}{2^{k_j}}\sum^{2^{k_j}}_{m_j=1}\mathbb{P}[ \widehat{M}_j\neq m_j\vert m_j~\text{is~sent} ]; 
    \end{align}
    \item The information leakage about the messages  $S_1$ and $S_2$ at the eavesdropper $\mathbf{L}_e\triangleq {I}(S_1S_2;Z^n)$.
\end{enumerate}

\begin{definition}
A $((k_l)_{l\in \mathcal{L}}, n, (P_l)_{l\in \mathcal{L}})$ code is said $((\epsilon_{S_i})_{i\in\mathcal{I}},(\epsilon_{M_j})_{j\in\mathcal{J}} )$-reliable if $\mathbf{P}^{(S_i)}_e\leq \epsilon_{S_i}$, $i\in\mathcal{I}$, and $\mathbf{P}^{(M_j)}_e\leq \epsilon_{M_j}$, $j\in\mathcal{J}$, and $\delta$-secure if $\mathbf{L_e}\leq \delta$. Moreover, a rate tuple  $\big(\frac{k_l}{n}\big)_{l\in\mathcal{L}}$ is said to be $((\epsilon_{S_i})_{i\in\mathcal{I}},(\epsilon_{M_j})_{j\in\mathcal{J}} )$-achievable with power constraints $(P_l)_{l\in\mathcal{L}}$ if there exists an $((\epsilon_{S_i})_{i\in\mathcal{I}},(\epsilon_{M_j})_{j\in\mathcal{J}} )$-reliable and $\delta$-secure $((k_l)_{l\in \mathcal{L}}, n, (P_l)_{l\in \mathcal{L}})$ code.
  \end{definition}
 \subsection{Code design} \label{mac_h_cd}
The design of the reliability layer is similar to that discussed in Section~\ref{rellyr_arch2}, which is detailed in Algorithm \ref{alg2}. Define $\Lambda_1\triangleq\{0,1\}^{q_1}\backslash \{\mathbf{0}\}$ and $\Lambda_2\triangleq\{0,1\}^{q_2}\backslash \{\mathbf{0}\}$ . For the design of the \textcolor{black}{security layer}, we need to design functions $(\varphi_{\lambda_1},  \varphi_{\lambda_2})$ and $(\psi_{\lambda_1}, \psi_{\lambda_2})$, where $\lambda_1\in\Lambda_1$ and $\lambda_2\in \Lambda_2$ are specific seeds,  similar to Section~\ref{SL}, such that the information leaked about secret messages is small, i.e., $I(S_1S_2;Z^n)\leq \delta$, for some $\delta>0$.
\subsection{Simulations}\label{mac_h_sim}
\subsubsection{Average probability of error}
We consider the channel model \eqref{mod1aarc3} with {\color{black}$\sigma^2_Y=1$,  $h_1=1$, $h_j=1$ for  $j\in \mathcal{J}$, and $L\in\{2,4\}$}. Set the transmitter message length $q_i\in\{4,6,8,10\}$,  $i\in\mathcal{I}$, helper  message length $q_j\in\{4,6,8,10\}$, $j\in \mathcal{J}$, and $n\in\{12,16,20,24\}$. For the design of the reliability layer for $i\in \mathcal{I}$ and $j\in \mathcal{J}$,  the autoencoders are trained for $(n, q_i, q_j)=(12,4, 4)$, $(n, q_i, q_j)=(16, 6,6)$, $(n, 
 q_i, q_j)=(20, 8, 8)$, and $(n, q_i, q_j)=(24,10, 10)$  over $N_e=600$ with $2\cdot 10^5$ random input messages at $l_r=0.0001$ with a  $b_s=500$.
\begin{figure}[h]
    \centering
    \includegraphics[trim=1.2cm 6.5cm 2.5cm 7cm,clip,width=0.45\textwidth]{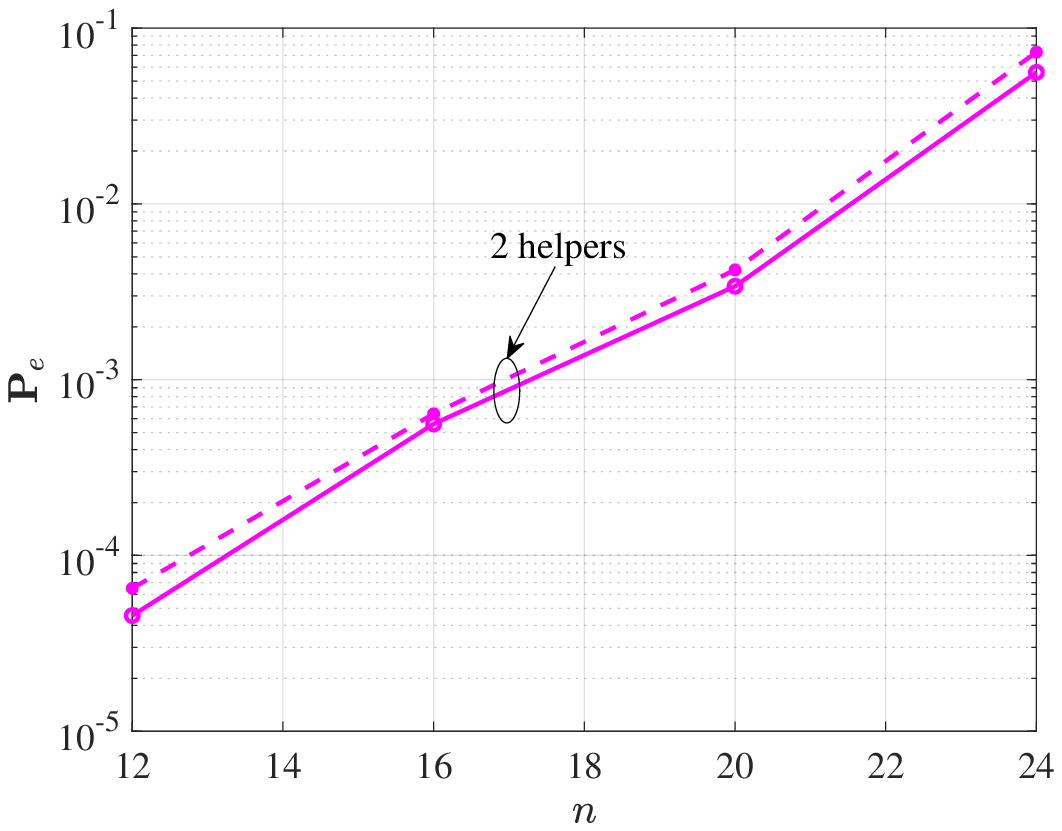}
    \caption{Average probability of error $\mathbf{P}_e^{(M_j)}$  versus blocklength. $\sigma^2_Y=1$, $P_1=4$~, $P_2=6$, and ~$P_j=16$, $j\in\{3,4\}$. The channel gains of the transmitter are $h_1=1$, $h_2=1$, and $h_j=1$, $j\in\{3,4\}$.}
    \label{fig:error_helper_2secret}
\end{figure}

\begin{figure}[h]
    \centering
    \includegraphics[trim=1.2cm 6.5cm 2.5cm 7cm,clip,width=0.45\textwidth]{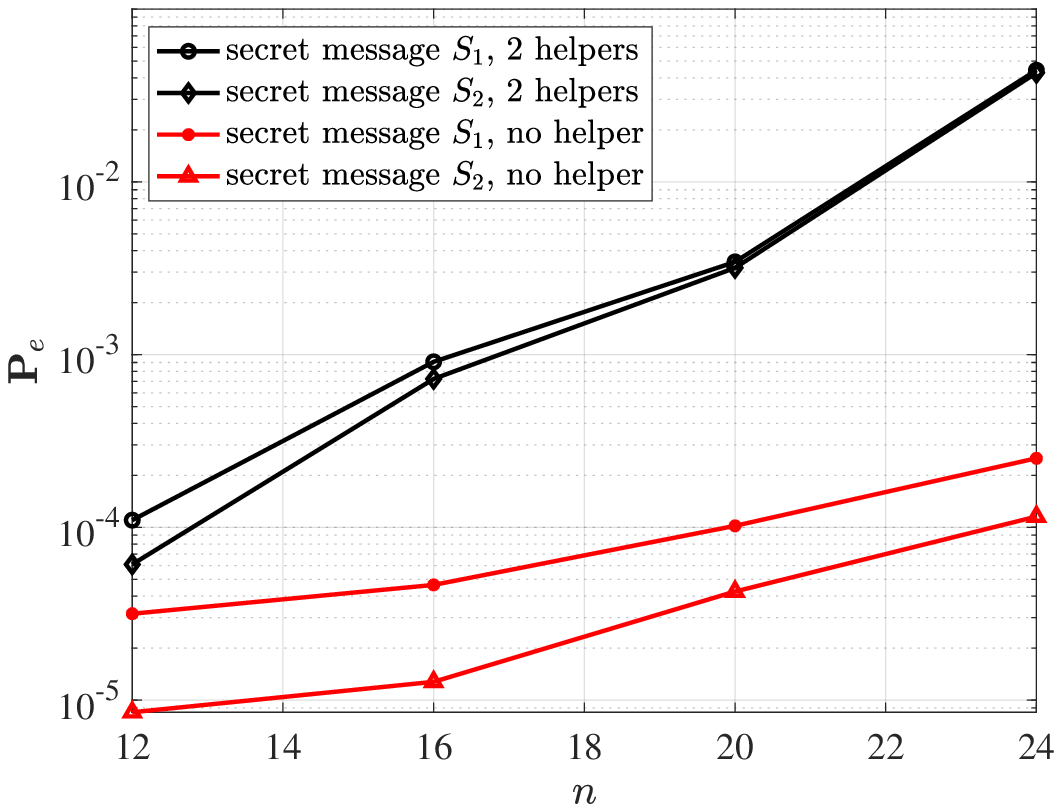}
    \caption{Average probability of error $\mathbf{P}_e^{(S_i)}$  versus blocklength  when $k_1=1$ and $k_2=1$. $\sigma^2_Y=1$, $\sigma^2_Z=1$, $P_1=4$~, $P_2=6$, and ~$P_j=16$, $j\in\{3,4\}$. The channel gains of the transmitter are $h_1=1$, $h_2=1$, and $h_j=1$, $j\in\{3,4\}$.}
    \label{fig:error_transmitters_2secret}
\end{figure}
\begin{figure}[h]
    \centering
    \includegraphics[width=0.45\textwidth]{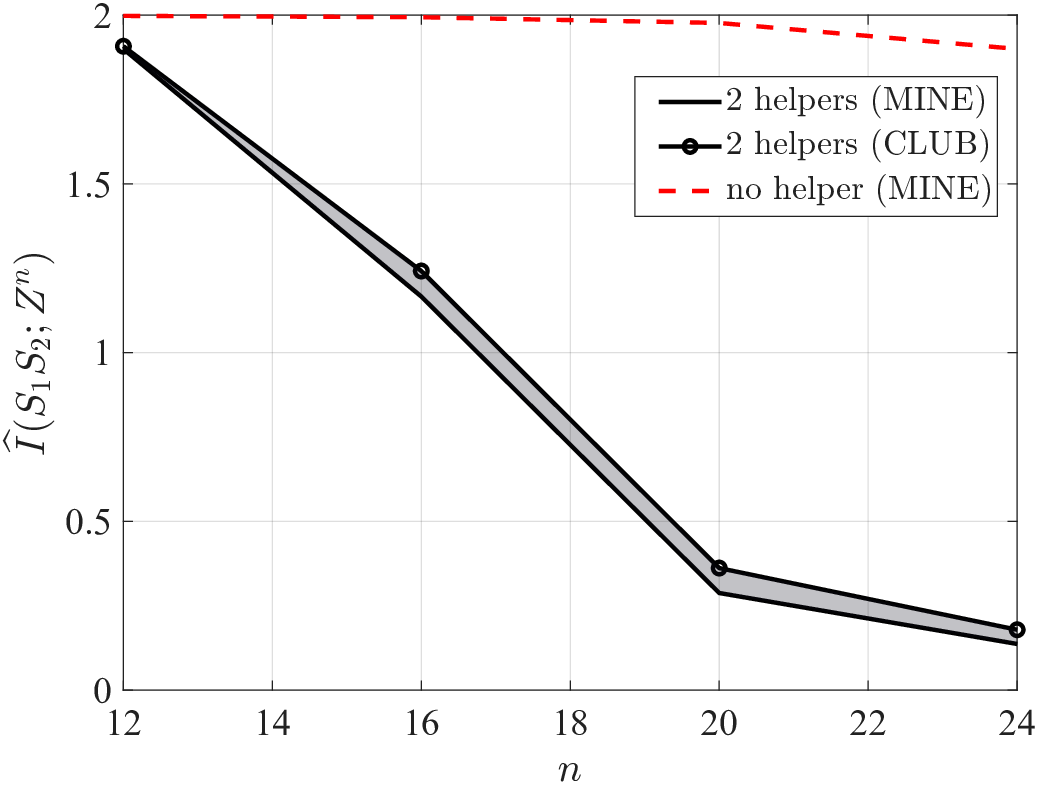}
    \caption{Information leakage versus blocklength obtained for $g_1=1$, $g_2=1$, and $g_j=0.3$, $j\in\{3,4\}$ when $k_1=1$ and $k_2=1$. $\sigma^2_Y=1$, $\sigma^2_Z=1$, $P_1=4$~, $P_2=6$ and ~$P_j=16$, $j\in\{3,4\}$. 
    }
    \label{fig:leakage_multihelpers}
\end{figure}

\textit{Testing}:
 To evaluate the average probability of error $\mathbf{P}_e^{(M_j)}$, $j\in\mathcal{J}$, we first generate the inputs $V_i\in\{0,1\}^{q_i}$, $i\in\mathcal{I}$, and $M_j\in\{0,1\}^{q_j}$, $j \in \mathcal{J}$. Then, $V_i$  is passed through the trained encoder $e^0_{i,n}$ and $M_j$ is passed through the trained encoder  $e^0_{j,n}$, which generates the codewords $X_i^n$ and $X_j^n$, respectively, and the channel output 
 \begin{align*}
     Y^n\triangleq \sum_{i=1}^{2}\sqrt{h_i}X^n_i+\sum_{j=3}^{L}\sqrt{h_j}X^n_j+N^n_Y.
 \end{align*}
 Finally, the trained decoder $d^0_{j,n}$ forms an estimate of~$M_j$, $\text{for}~j=L~\text{to}~3,$ from 
 \begin{align*}
     Y^n-\sum_{j'=j+1}^{L}\sqrt{h_{j'}}\widehat{X}^n_{j'},
 \end{align*}
 as described in Section \ref{rellyr_arch2}.

Consider $\varphi_i$ and $\psi_i$, $i\in\mathcal{I}$, with the transmitter message length $q_i\in\{4,6,8,10\}$. We chose the seeds as $\lambda_i\in \{0001, 000001, 00000001, 0000000001\}$ for the different values of $q_i$.  Set the secret length $k_i = 1$, $i\in\mathcal{I}$. To evaluate the average probability of error $\mathbf{P}_e^{(S_i)}$, $i\in\mathcal{I}$, for the secret message, the trained encoder $e^0_{i,n}$  encodes the messages $S_i \in \{0,1\}^{k_i}$  as $e_{i,n}^0(\varphi_i(S_i,B_i))$, $i\in\mathcal{I}$, as described in Section~\ref{SL}, where $B_i \in \{0,1\}^{q_i-k_i}$ is a sequence of $q_i-k_i$ bits generated uniformly at random. The trained decoder $d^0_{2,n}$ forms 
\begin{align*}
    \widehat{S}_2 \triangleq \psi_2(d^0_{2,n}(Y^n-\textstyle\sum_{j=3}^L\sqrt{h_j}\widehat{X}_j^n)),
\end{align*}
and the trained decoder $d^0_{1,n}$ forms 
\begin{align*}
    \widehat{S}_1 \triangleq \psi_1(d^0_{1,n}(Y^n-\sqrt{h_2}\widehat{X}_2^n-\textstyle\sum_{j=3}^L\sqrt{h_j}\widehat{X}_j^n)),
\end{align*}
as described in Section~\ref{SL}. 

Figure~\ref{fig:error_helper_2secret} shows the probability of error $\mathbf{P}^{(M_j)}_e$, $j\in\mathcal{J}$, for the unprotected messages. 
 Figure \ref{fig:error_transmitters_2secret} shows the average probability of error for the secret message $\mathbf{P}_e^{(S_i)}$, $i\in\mathcal{I}$. As
expected, similar to Section \ref{sec:error_arch2}, we observe the probability of error increases as the number of helpers increases. {\color{black}From Figure \ref{fig:error_transmitters_2secret}, we also observe that   $\mathbf{P}_e^{(S_1)}>\mathbf{P}_e^{(S_2)}$, since $P_2>P_1$.}
\subsubsection{Information leakage evaluation}
For the simulations, we consider the model in \eqref{mod1barc3} with $\sigma_Z^2=1$ and secret length $k_i=1$, $i\in\mathcal{I}$. {\color{black} We have $L\in\{2,4\}$, $g_1=1$, $g_j=0.3$, helper  message length $q_j\in\{4,6,8,10\}$,  $j\in \mathcal{J}$, and $n\in\{12,16,20,24\}$.} Consider $\varphi_i$ and $\psi_i$ with the transmitter message length $q_i\in\{4,6,8,10\}$, $i\in\mathcal{I}$. We chose the seeds as $\lambda_i\in \{0001, 000001, 00000001, 0000000001\}$ for the different values of $q_i$. We compute  $I(S_1S_2;Z^{n})$, similar to \textcolor{black}{Sections \ref{mine} and \ref{club}} in Figure~\ref{fig:leakage_multihelpers} for different number of helpers. As
expected, similar to Section \ref{sec:lk_arch2}, we observe that the information leakage decreases as the number of helpers increases.

\subsubsection{Discussion}
Figure~\ref{fig:leakage_multihelpers} shows there is an improvement in information leakage for codes with a helper compared to codes without a helper for the Gaussian multiple access wiretap channel.

Figures \ref{fig:error_helper_2secret}, \ref{fig:error_transmitters_2secret}, and \ref{fig:leakage_multihelpers} show that we designed a code with rates  $\Big(\frac{1}{20},\frac{1}{20}, \frac{8}{20}, \frac{8}{20}\Big)$ that is $(\epsilon_{S_1}= 3.4\cdot 10^{-3},\epsilon_{S_2}= 3.1\cdot 10^{-3},  \epsilon_{M_3}=3.4\cdot 10^{-3}, \epsilon_{M_4}=4.2\cdot 10^{-3}, \textcolor{black}{\delta=3.6 \cdot 10^{-1} })$- achievable with power constraint $(4, 6, 16, 16)$.

 \section{Concluding Remarks}\label{cr}

For the first time in the literature, we designed explicit and short blocklength codes for the Gaussian wiretap channel in the presence of a helper that cooperates with the transmitter.  Our proposed codes showed significant improvement in information leakage compared to existing point-to-point codes, even when the transmitter has adverse channel conditions, i.e., the eavesdropper experiences less channel noise than the legitimate receiver. We proposed a framework that separates the code design into two layers: a reliability layer and a \textcolor{black}{security layer}. We implemented the reliability layer with an autoencoder based on SIC and the \textcolor{black}{security layer} with universal hash functions. We further proposed an alternative  autoencoder architecture in which the decoders independently estimate messages without successively canceling interference by the receiver during training. We found that the training time is significantly reduced compared to the model where interference is successively canceled. We also showed that our code design is applicable to settings where multiple transmitters send secret messages, e.g., Gaussian multiple access wiretap channels.

 \textcolor{black}{While in our setting, both intended receiver and eavesdropper have AWGN channels with constant gains, if the eavesdropper has a fast Rayleigh fading channel with additive white Gaussian noise, as in \cite{zang2010}, then  a positive secrecy rate is achievable with
artificial noise and bursting strategy even when Bob’s channel
is  worse than Eve’s average channel gain \cite{zang2010}. Additionally, our scenario does not involve multiple transmit antennas or amplifying relays \cite{negi2005}.
Designing short blocklength codes for the framework in \cite{zang2010, negi2005}  under these constraints remains an open problem and presents an interesting direction for future research.}
\textcolor{black}{Another open problem is to design codes with larger blocklengths.} We note that  there exist autoencoder-based channel codes, such as CNN-autoencoders \cite{CNN-AE} and Turbo-autoencoders \cite{jiang2019turbo}, with blocklengths up to $n \leq 1000$. However, incorporating these constructions into the
reliability layer of our framework and evaluating their security guarantees remains
an open problem. \textcolor{black}{In particular, a main bottleneck is the computational complexity of MINE and CLUB to
evaluate the information leakage for larger blocklengths as the total number of parameters increases with the size of the input layer.}

\bibliographystyle{IEEEtran}
\bibliography{mac}
\end{document}